\documentclass[a4paper,10pt,twocolumn]{article}
\usepackage{graphicx}
\usepackage[top=1.78cm, bottom=2.38cm, left=1.78cm, right=1.43cm]{geometry}
\usepackage{sectsty}
\usepackage{amsmath}
\usepackage{algorithm}
\usepackage[noend]{algpseudocode}
\usepackage[utf8x]{inputenc}
\makeatletter
\def\BState{\State\hskip-\ALG@thistlm}
\makeatother
\usepackage{amssymb} 
\usepackage{blindtext}
\makeatletter
\renewcommand{\paragraph}{\@startsection{paragraph}{20}{12pt}{4.0ex plus 1ex minus .2ex}{1.5ex plus .2ex}
   {\normalfont\normalsize\bfseries}}
\makeatother
\stepcounter{secnumdepth}
\stepcounter{tocdepth}
\makeatletter
\def\@seccntformat#1{\@ifundefined{#1@cntformat}
   {\csname the#1\endcsname\space}
   {\csname #1@cntformat\endcsname}}
\newcommand\section@cntformat{\thesection.\space}       
\makeatother
\usepackage{setspace}
\usepackage{titlesec}
\titlespacing*{\section}
{0pt}{6.0pt plus 0pt minus 0pt}{6pt plus 0pt}
\titlespacing*{\subsection}
{12pt}{6pt plus 6pt  minus 6pt}{3pt plus 6pt  minus 6pt}
\titlespacing*{\subsubsection}
{12pt}{6pt plus 6pt  minus 6pt}{3pt plus 6pt  minus 6pt}
\titlespacing*{\paragraph}
{12pt}{6pt plus 6pt  minus 6pt}{3pt plus 6pt  minus 6pt}
\titleformat{\subsection}
       {\normalfont\fontsize{10}{10}\bfseries}{\thesubsection}{1em}{} 
\titleformat{\subsubsection}
       {\normalfont\fontsize{10}{10}\bfseries\itshape}{\thesubsubsection}{1em}{} 
\titleformat{\paragraph}
       {\normalfont\fontsize{10}{10}\bfseries\itshape}{\theparagraph}{1em}{} 
 \usepackage[numbers]{natbib} 
\usepackage{multirow}
\usepackage[table,xcdraw]{xcolor}
\usepackage{lscape}
\usepackage{booktabs}
\usepackage{hhline}
\usepackage{url}

\usepackage{usebib}
\def\authorrefmark#1{\textsuperscript{#1}}
\usepackage{tikz}
\sectionfont{\large}
\makeatletter
\newcommand\notsosmall{\@setfontsize\notsosmall{8.99}{9}}
\newcommand\notsosmallnew{\@setfontsize\notsosmallnew{9.99}{10}}
\newcommand\largefont{\@setfontsize\largefont{23.99}{24}} 
\newcommand\smallfont{\@setfontsize\smallfont{10.99}{11}}   
\makeatother
\usepackage{fancyhdr}
\fancypagestyle{plain}{
  \fancyhead[R]{\notsosmallnew \thepage}  
}  
\usepackage[labelfont=bf]{caption}
\DeclareCaptionLabelSeparator{dot}{. }
\usepackage[labelsep=dot]{caption}
\usepackage{blindtext}
\usepackage{caption} 
\usepackage{titling}
\usepackage{enumitem}
\begin{document}
\title {\largefont Efficient and Secure ECDSA Algorithm and its Applications: A Survey \vspace{-15pt}}
\author{\smallfont{Mishall Al-Zubaidie\authorrefmark{1,2}, Zhongwei Zhang\authorrefmark{2} and Ji Zhang\authorrefmark{2}\vspace{.1pt}}\and
{\notsosmall \authorrefmark{1}Thi-Qar University, Nasiriya 64001, Iraq}\\
{\notsosmall \authorrefmark{2} Faculty of Health, Engineering and Sciences,}\\ {\notsosmall University of Southern Queensland, Toowoomba, QLD 4350, Australia\vspace{-20pt}} 
}
\date{}
\maketitle
\begin{notsosmall}
\noindent{\textbf{\emph{\ \ \  Abstract:}}} Public-key cryptography algorithms, especially elliptic curve cryptography (ECC) and elliptic curve digital signature algorithm (ECDSA) have been attracting attention from many researchers in different institutions because these algorithms provide security and high performance when being used in many areas such as electronic-healthcare, electronic-banking, electronic-commerce, electronic-vehicular, and electronic-governance. These algorithms heighten security against various attacks and the same time improve performance to obtain efficiencies (time, memory, reduced computation complexity, and energy saving) in an environment of constrained source and large systems. This paper presents detailed and a comprehensive survey of an update of the ECDSA algorithm in terms of performance, security, and applications.\\

\end{notsosmall}
\setlength{\abovedisplayskip}{3pt}
\setlength{\belowdisplayskip}{3pt}
\section{Introduction}
Public key encryption algorithms such as elliptic curve cryptography (ECC) and elliptic curve digital signature algorithm (ECDSA) have been used extensively in many applications \cite{p3} especially in constrained-resource environments due to the effectiveness of their use it in these environments. These algorithms have appropriate efficiency and security for these environments.
Constrained resource environments such as wireless sensor network (WSN), radio frequency identifier (RFID), and smart card require high-speed, low consumption ability, and less bandwidth. ECC has been considered to be appropriate for these constrained-source environments \cite{p70}. These algorithms provide important security properties. For example, ECDSA provides integrity, authentication, and non-repudiation. ECDSA has been proven to be efficient in its performance because it uses small keys; thus the cost of computation is small compared with other public key cryptography algorithms, such as Rivest Shamir Adleman (RSA), traditional digital signature algorithm (DSA) and ElGamal. For example, ECDSA with a 256-bit key offers the same level of security for the RSA algorithm with a 3072-bit key \cite{p61,p27}. Table~\ref{survey:tab1} \cite{p99,p100,p5} shows a comparison of key sizes for public key signature algorithms.
\begin{table*}[!t]
\tiny
\centering
\caption{Keys sizes and some information for public key algorithms}
\label{survey:tab1}
\resizebox{\textwidth}{!}{%
\begin{tabular}{|l|l|l|l|l|l|l|l|l|l|l|}
\hline
\rowcolor[HTML]{EFEFEF} 
\hline
Algorithm & \multicolumn{5}{l|}{\cellcolor[HTML]{EFEFEF}Keys sizes} & Ratio & Author(s) & Year & \begin{tabular}[c]{@{}l@{}}Mathmatical\\ problem\end{tabular} & \begin{tabular}[c]{@{}l@{}}Other \\ algorithms\end{tabular} \\ \hline
RSA &  &  &  &  &  &  & \begin{tabular}[c]{@{}l@{}}Rivest,\\  Shamirand, \\ and Adleman\end{tabular} & 1978 & \begin{tabular}[c]{@{}l@{}}Integer\\ factorization\end{tabular} & Rabin \\ \cline{1-1} \cline{8-11} 
Elgamal  &  &  &  &  &  &  &  Taher Elgamal & 1985 &  &  \\ \cline{1-1} \cline{8-9}
DSA & \multirow{-3}{*}{1024} & \multirow{-3}{*}{2048} & \multirow{-3}{*}{3072} & \multirow{-3}{*}{7680} & \multirow{-3}{*}{15360} &  & \begin{tabular}[c]{@{}l@{}}David W. \\ Kravitz\end{tabular} & 1991 & \multirow{-2}{*}{\begin{tabular}[c]{@{}l@{}}Multiplicative\\ group\end{tabular}} & \multirow{-2}{*}{\begin{tabular}[c]{@{}l@{}}Schnorr,\\ Nyberg-Rueppel\\ \end{tabular}} \\ \cline{1-6} \cline{8-11} 
ECDSA & 160-223 & 224-255 & 256-383 & 384-511 & 512-more & \multirow{-4}{*}{1:6-30} & Scott Vanstone & 1992 & \begin{tabular}[c]{@{}l@{}}Elliptic curve\\ discrete log.\\ (ECDLP)\end{tabular} & ECIES \\ \hline
\end{tabular}}
\end{table*}
\\The preservation of efficiency and security in the ECDSA is important. On one hand, several approaches have been developed to improve the efficiency of the ECDSA algorithm to reduce the cost of computation, energy, memory, and consumption of processor capabilities. The operation that consumes more time in ECC/ECDSA is the point multiplication (PM) or scalar multiplication (SM). ECC is used PM for encryption and decryption, while ECDSA is used this operation to generate and verify the signature \cite{p85}. One can improve the PM efficiency by improving finite field arithmetic (such as inversion, multiplication, and squaring), elliptic curve model (such as Hessian and Weierstrass), point representation (such as Projective and Jacobian), and the methods of PM (such as Comb and Window method) \cite{p62}. Many researchers have made improvements to the PM to increase the performance of the ECC/ECDSA as we will see in Section 4.
\\On the other hand, the security improvement in ECDSA is no less important than its efficiency because this algorithm is designed primarily for the application of security properties. ECDSA, like previous algorithms, may possibly suffer from some of the security vulnerabilities such as random weak, bad random source \cite{p3} or leaking bits of the private key. Also, many researchers have made improvements to close security gaps in the ECC/ECDSA algorithm by providing countermeasures against various attacks. But when selecting countermeasure there should be a balance between security and efficiency \cite{p92}. To maintain the security of these algorithms it is important to use finite fields (either prime or binary) recommended by credible institutions such as the Federal Information Processing Standard (FIPS) or National Institute of Standards and Technology (NIST). The choice of appropriate curves and finite fields according to the authoritative organizations' standards leads to secure ECDSA's implementations \cite{p63}. Therefore, we note from the above that any encryption algorithm or signature should possess a high performance and security level.
\titlespacing*{\section}
{0pt}{6.0pt plus 0pt minus 0pt}{6pt plus 0pt}
\titlespacing*{\subsection}
{12pt}{6pt}{3pt}
\titlespacing*{\subsubsection}
{12pt}{6pt}{3pt}
\titlespacing*{\paragraph}
{12pt}{6pt}{3pt}
\subsection{Our Contributions}
Our contribution in this survey is to provide an updated study of three important parts in ECC/ECDSA. These items can be summarized as follows:
\begin{itemize}[noitemsep,nolistsep]
\item Presenting a study on the efficiency of ECC/ECDSA in terms of speed, time, memory, and energy.
\item Investigating the security problems of the ECC/ECDSA and countermeasures. 
\item Giving a description of the most important applications that use ECC/ECDSA algorithms. 
\end{itemize}
\titlespacing*{\section}
{0pt}{6.0pt plus 0pt minus 0pt}{6pt plus 0pt}
\titlespacing*{\subsection}
{12pt}{6pt plus 6pt  minus 6pt}{3pt plus 6pt  minus 6pt}
\titlespacing*{\subsubsection}
{12pt}{6pt plus 6pt  minus 6pt}{3pt plus 6pt  minus 6pt}
\titlespacing*{\paragraph}
{12pt}{6pt plus 6pt  minus 6pt}{3pt plus 6pt  minus 6pt}
\subsection{Structure of the paper}
The remainder of this paper is structured as follows: Section 2 provides basic concepts and general information about ECC and ECDSA algorithms. Section 3 describes existing surveys about ECC/ECDSA. Efficiency improvement on ECDSA algorithm is presented in Section 4. In Section 5, we will show the security improvement on ECDSA algorithm through using countermeasures against attacks. The ECDSA applications are described in Section 6. Finally, we will present the conclusion and future work on this survey in Section 7.
\titlespacing*{\section}
{0pt}{6.0pt plus 0pt minus 0pt}{6pt plus 0pt}
\titlespacing*{\subsection}
{12pt}{6pt plus 6pt  minus 6pt}{3pt plus 6pt  minus 6pt}
\titlespacing*{\subsubsection}
{12pt}{6pt plus 6pt  minus 6pt}{3pt plus 6pt  minus 6pt}
\titlespacing*{\paragraph}
{12pt}{6pt plus 6pt  minus 6pt}{3pt plus 6pt  minus 6pt}
\section{Preliminaries of ECDSA}
In this section, we will present fundamentals about elliptic curve cryptography (ECC) and basic concepts of the ECDSA algorithms.
\titlespacing*{\section}
{0pt}{6.0pt plus 0pt minus 0pt}{6pt plus 0pt}
\titlespacing*{\subsection} {12pt}{-0.1em}{-0.4em}
\titlespacing*{\subsubsection}
{12pt}{6pt plus 6pt  minus 6pt}{3pt plus 6pt  minus 6pt}
\titlespacing*{\paragraph}
{12pt}{6pt plus 6pt  minus 6pt}{3pt plus 6pt  minus 6pt}
 \subsection{ECC}
\titlespacing*{\section}
{0pt}{6.0pt plus 0pt minus 0pt}{6pt plus 0pt}
\titlespacing*{\subsection}
{12pt}{6pt plus 6pt  minus 6pt}{3pt plus 6pt  minus 6pt}
\titlespacing*{\subsubsection}
{12pt}{6pt plus 6pt  minus 6pt}{3pt plus 6pt  minus 6pt}
\titlespacing*{\paragraph}
{12pt}{6pt plus 6pt  minus 6pt}{3pt plus 6pt  minus 6pt}
ECC has been used to encrypt data to provide confidentiality in the communications network with limited capacity in terms of power and processing. This algorithm was independently proposed by Neal Koblitz and Victor Miller in 1985 \cite{p14}. It depends on the discrete logarithm problem (DLP) which is impervious to different attacks when selecting parameters accurately \cite{p18}, i.e difficulty obtaining \textit{k} from \textit{P} and \textit{Q} ( where \textit{k} is integer and, \textit{P} and \textit{Q} are two points on the curve). Small parameters used in ECC help to perform computations quickly. These computations are important in constrained-source environments that require processing power, memory, bandwidth or power consumption \cite{p5}. ECC provides encryption, signature and keys exchange approaches \cite{p14}. Many operations are performed in ECC algorithms (shown in four layers) as shown in Figure 1 \cite{p23}.
\tikzstyle{block} = [rectangle, draw, fill=white!20, text width=9em, text centered, minimum height=3em ,minimum width=4em] 
\tikzstyle{block1} = [rectangle, draw, fill=white!20, text width=5em, text centered, minimum height=3em ,minimum width=4em]
\tikzstyle{block2} = [rectangle, draw, fill=white!20, text width=7em, text centered, minimum height=3em ,minimum width=4em]
\tikzstyle{block3} = [rectangle, draw, fill=white!20, text width=3em, text centered, minimum height=2em ,minimum width=4em]
\tikzstyle{line} = [draw, -latex']   
\tikzstyle{cloud} = [draw, ellipse,fill=white!20, node distance=3cm, minimum height=2em]
\usetikzlibrary{arrows}
\begin{figure*}[t]
\centering
\begin{tikzpicture}
\scriptsize
\node [block] at (-12,5)(rec1) {\textbf{The first layer \\cryptographic protocol}};
\node [block] at (-12,3)(rec2) {\textbf{The second layer\\ elliptic curve point multiplication}};
\node [block] at (-12,1)(rec3) {\textbf{The third layer\\ point operations}};
\node [block] at (-12,-1)(rec4) {\textbf{The lower finite field arithmetic}};
\node [block] (rec5) at (-3.4,5) {\textbf{ECC (ECIES, ECDSA and ECDH)}};
\node [block] (rec6) at (-3.4,3) {\textbf{Point multiplication (PM)}};
\node [block] (rec7) at (-5,1) {\textbf{Point additioin}};
\node [block] (rec8) at (-2,1) {\textbf{Point doubling}};
\node [block2] at (-6.4,-1)(rec9) {\textbf{Multiplication}};
\node [block1] at (-4.2,-1)(rec10) {\textbf{Addition}};
\node [block1] at (-2.3,-1)(rec11) {\textbf{Squaring}};
\node [block1] at (-0.4,-1)(rec12) {\textbf{Inversion}};
\path [line] (-10.0,5.0) -- (-8.0,5.0);
\path [line] (-10.0,3.0) -- (-8.0,3.0);
\path [line] (-10.0,1.0) -- (-8.0,1.0);
\path [line] (-10.0,-1.0) -- (-8.0,-1.0);
\draw (rec5) --(rec6);
\draw (-3.4,2.55) --(rec7);
\draw (-3.4,2.55) --(rec8);
\draw (-5,.55) ->(-6.5,-0.54);
\draw (-5,.55) --(-4.0,-0.54);
\draw (-5,.55) --(-2.4,-0.54);
\draw (-5,.55) --(-0.4,-0.54);
 \draw (-2.0,.55) --(-6.5,-0.54);
\draw (-2.0,.55) --(-4.0,-0.54);
\draw (-2.0,.55) --(-2.4,-0.54);
\draw (-2.0,.55) --(-0.4,-0.54);
\draw [-triangle 60] (-13.5,5.7) rectangle (0.5,-1.5);
\end{tikzpicture}
 \caption{Arithmetic operations in ECC hierarchy}
  \label{fig3:arith_ecc}
\end{figure*}
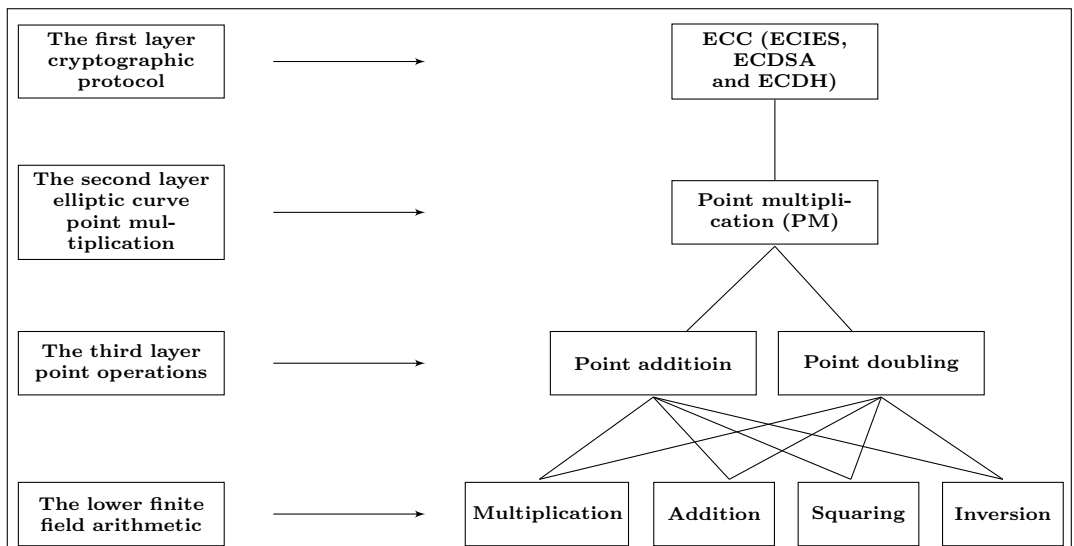
ECC uses two finite fields (prime field and binary field). The binary field uses two types to represent basis ( normal and polynomial basis) \cite{p5}, and is well suited to implementation in hardware \cite{p23}. Let $F_q$ indicates field type, if \textit{q}=\textit{p} (where $p>3$) then ECC uses prime field ($F_p$). In the second case, if \textit{q}=2 then ECC uses binary field ($F_{2^m}$) where m is the prime integer \cite{p67}. ECC consists of a set of points ($x_i,y_i$), where $x_i ,y_i$  are integers and the point at infinity (O) provides an identity for Abelian group rule that satisfies long form for Weierstrass equation (with Affine coordinate).
\begin{equation}
y^2 + a_1xy + a_3y = x^3 + a_2x^2 + a_4x + a_6
\end{equation}
When prime field is used over ECC, the simplified equation is as follows:{\color{white} highlighted}
\begin{equation}
y^2 = x^3 + ax + b 
\end{equation}
Where $a,b \in F_p, 4a^3+27b^2 \neq 0(mod p)$. The law of chord-and-tangent is used in ECC to add two points on the curve. Let us suppose that \textit{P} and \textit{Q} are two points on the curve; these two points have coordinates $(x_1,y_1), (x_2,y_2)$ respectively and the sum of these two points is equal to a new point $R(x_3,y_3)$ (i.e $P(x_1,y_1)$+$Q(x_2,y_2)$=$R(x_3,y_3)$) \cite{p5}. ECC uses two operations for addition that are point addition($P$+$Q$) and point doubling ($P$+$P$) (Figure 1), as in the following equations:\\
In the case of the point addition $(P+Q)$ where $P$ and $Q$ $\in E(F_p)$:\\
with using the slope $\lambda= \frac{y_2-y_1}{x_2-x_1}$
\begin{equation}
x_3= \lambda^2-x_1-x_2,\  y_3=\lambda^2(x_1-x_3)-y_1
\end{equation}
In the case of the point doubling $(P+P)$ where $P \in E(F_p)$:\\
with using slope $\lambda= \frac{3x_1^2+a}{2y_1}$
\begin{equation}
x_3=\lambda^2-2x_1,\ y_3=\lambda^2(x_1-x_3)-y_1
\end{equation}
When the binary field is used over ECC, the simplified equation is as follows:
\begin{equation}
y^2+xy = x^3 + ax^2 + b
\end{equation}
Where $a,b \in F_{2^m}, b \neq 0$ \cite{p5}, as addition operations are in the following form:\\
In the case point addition $(P+Q)$ where $P$ and $Q \in E(F_{2^m})$:\\
with using the slope $\lambda= \frac{y_1+y_2}{x_1+x_2}$
\begin{equation}
x_3=\lambda^2+ \lambda +x_1+x_2+a,\ y_3=\lambda(x_1+x_3)+x_3+y_1
\end{equation}
In the case point doubling $(P+P)$ where $P \in E(F_{2^m})$:\\
\begin{equation}
x_3= x_1^2+\frac{b}{x_1^2},\ y_3= x_1^2+(x_1+\frac{y_1}{x_1})x_3+x_3
\end{equation}
ECC operations for encryption and decryption are explained through the algorithm~\ref{alg:algorithm1} \cite{p12}.
\begin{algorithm}[t]
\caption{ECC encryption and decryption algorithm:} 
\label{alg:algorithm1}
\begin{algorithmic}[1]
\BState Alice and Bob use same parameters domain $D=\{a,b,q,G,n,h\}$, where $a,b$ are coefficient, $q$ is field type, $G$ is base point, $n$ is order point and $h$ is cofactor.
\BState Alice selects random integer $d_A$ as private key.
\BState Alice generates public key $Q=d_AG$ and sends $Q$ and $G$ to Bob. 
\BState Bob receives message $m$ from Alice, selects random integer as private key $d_B$ where $d_B\le n$.
\BState Bob encrypts message $m$ with point $P$ in elliptic curve $E(F_q)$.
\BState Bob computes $C_1=P+d_BQ$, $C_2=d_BG$ and sends $C_1$ and $C_2$ to Alice.
\BState Alice receives Bob's message and decrypts the message by computing $C_1-d_AC_2$ to obtain plaintext.
\end{algorithmic}
\end{algorithm}
\subsection{ECDSA}
ECDSA algorithm is used to warrant data integrity to prevent tampering with the data. This algorithm was proposed by Scott Vanstone in 1992. Data integrity of the message is important in the networks because the attacker can modify the message when it is transferred from source to destination \cite{p29}. Many organizations use it as standard such as ISO (1998), ANSI (1999) and, IEEE and NIST (2000) \cite {p5}. This algorithm is similar to the digital signature algorithm (DSA), where both algorithms depend on the discrete logarithm problem (DPL), but ECDSA algorithm uses a set of points on the curve and the generating keys are small. ECDSA algorithm with key length 160-bit provides the equivalent for symmetric cryptography with a key length of 80-bit \cite{p29}. It is dramatically convenient for devices with constrained-source because it uses small keys and provides computation speed in the signature \cite{p47}. Moreover, four point multiplication operations used in ECDSA algorithm: one is in public key generation, one for signature generation and two for signature verification \cite{p39}. In addition, this algorithm consists of three procedures: key generation, signature generation, and signature verification. These operations are explained as follows:
\begin{itemize}[noitemsep,nolistsep]
\item \textbf{Key generation:}
\begin{enumerate}[noitemsep,nolistsep]
\item [1] Select a random or pseudorandom integer $d$ in the interval $[1,n-1]$.
\item [2] Compute $Q = dG$
\item [3] Public key is $Q$, private key is $d$. 
\end{enumerate}
\item \textbf{Signature generation:}
\begin{enumerate}[noitemsep,nolistsep]
\item [1] Select a random or pseudorandom integer $\textit{k}$, 1 $\leq \textit{k} \leq$ 1.
\item [2] Compute $kG=(x_1,y_1)$ and convert $x_1$ to an integer $\bar{x}_1$. 
\item [3] Compute $r=x_1$ mod $n$.  If $r=0$ then go to step 1. 
\item [4] Compute $k^{-1}$ mod $n$.
\item [5] Compute SHA-1($m$) and convert this bit string to an integer $e$.
\item [6] Compute $s$= $k^{-1}(e+dr)$mod $n$. If $s=0$ then go to step 1.
\item [7] Signature for the message m is $(r,s)$.
\end{enumerate}
\item \textbf{Signature verification:}
\begin{enumerate}[noitemsep,nolistsep]
\item [1] Verify that $r$ and $s$ are integers in the interval [1,$n$-1]. 
\item [2] Compute SHA-1($m$) and convert this bit string to an integer $e$.
\item [3] Compute $w=s^{-1}$ mod $n$.
\item [4] Compute $u_1=ew$ mod $n$ and $u_2=rw$ mod $n$.
\item [5] Compute $X=u_1G + u_2Q$.
\item [6] If $X= \theta$ then reject the signature. Otherwise, convert the $x$-coordinate $x_1$ of $X$ to an integer $\bar{x}_1$, and compute $v= \bar{x}_1$ mod $n$.
\item [7] Accept the signature if and only if $v=r$. 
\end{enumerate}
\end{itemize}
ECDSA algorithm becomes unsuitable for signing messages (integration) if used poorly and incorrectly. Validation of domain parameters is important to ensure strong security against different attacks. This algorithm becomes strong if the parameters are well validated \cite{p120,p22}. The authors' recommendations are to update the validation scheme.
\subsection{ECDSA Implementations on Constrained Environments}
The performance and security presented by ECDSA algorithm, make it suitable for use in several implementations on WSN, RFID, and smart card. Digital signatures in ECDSA have better efficiency in devices with constrained-resource than DSA and RSA. Many authors have pointed to the possibility of using ECDSA with environments constrained-resource (memory, energy, and CPU capability). In the following Subsections, we will explain using ECDSA with these implementations.
\subsubsection{Wireless Sensor Network (WSN)} 
A WSN consists of a group of nodes that communicate wirelessly with each other to gather information about a particular environment in various applications \cite{p170}. This network often has restricted sources such as energy and memory. Therefore, a WSN needs efficient algorithms to reduce the complexity of the computation in order to increase the length of the network lifetime. In addition, it requires a high level of security to prevent attacks. Several researchers have investigated the use of ECDSA algorithm in WSN and explained that ECDSA is convenient for WSN.\\
For example, \citet{p45} presented a study in energy for the public key cryptography (ECC/ECDSA, RSA) on sensor node Mica2dot with Atmel ATmegal 128L (8bit). They found that transmission cost is double the receiving cost. They analysed signatures in ECDSA with a key of bit-160 and RSA with a key of bit-1024, signature verification cost in ECDSA is larger than a signature generation while RSA verification is smaller than a signature generation. They noted that ECDSA has less energy cost than RSA. They concluded that ECC/ECDSA is more effective than RSA and feasible in constrained-source devices (WSN) because it generates small keys and certificates with same security level in RSA. Also, implementation was applied for  160-bit ECC/ECDSA and 1024-bit RSA algorithms on sensor node MICAz \cite{p32}. The authors used hybrid multiplication to reduce access memory. ECDSA results on MICAz are signature generation=1.3s and signature verification = 2.8s. For the purpose of comparison, the authors implemented ECDSA also on TelosB, where MICAz results were slightly less than TelosB's results. The authors proved the possibility of using RSA and ECDSA on WSN. In addition, ECDSA (SHA-1) and RSA (AES) were analysed in several types of sensor nodes (Mica2dot, Mica2, Micaz, and TelosB) in terms of energy and time \cite{p9}.  ECDSA uses short keys (160-bit), which reduces memory, computation, energy and data size transmitted and thus is better than the RSA. \citet{p16} discussed the cost evaluation of energy (communication, computation) on the WSN (TelosB and MICAz) through the Kerberos key distribution protocol (symmetric encryption) and ECDH-ECDSA key agreement protocol (asymmetric encryption). They noted that TelosB sensor node consumes less power than MICAz and Kerberos performs better than ECDH-ECDSA. Moreover, \citet{p46} implemented ECDSA on Micaz motes with the binary field (163-bit). They improved signature verification via cooperation idea of the adjacent nodes. Also, ECDSA's implementation was presented in sensor node (IRIS) \cite{p10}, but because this node supports 8-bit of the microcontroller the author modified the SHA-1 code from 32 bits original to 8 bits. Through implementation, the original algorithm is better in size and time than a modified algorithm.  The author explained the possibility of using ECDSA algorithm with the sensor node (IRIS) held 8-bit microcontroller. \\
Recently, researchers in \cite{p93,p94} have applied the ECDSA algorithm as a light-weight authentication scheme in the WSN. This demonstrates the effectiveness and efficiency of using ECDSA in WSN in terms of security and performance. 
\subsubsection{Radio Frequency Identifier (RFID)}
Another implementation of ECDSA is Radio Frequency Identifier (RFID). RFID is one of the technologies used in wireless communication. This technology has been used significantly in various fields, but it suffers from constrained-resources such as area and power. Therefore, many researchers considered ECDSA algorithm to be the best choice in RFID tags because ECDSA is not required to store private/public keys as in symmetric cryptography algorithms.  \\
The hardware's implementation of ECDSA algorithm has been widely used on RFID technology \cite{p74}. Authors have used ECDSA algorithm to authenticate the entity. They have applied this algorithm to the prime field ($F_{p160}$) according to SECG standards. They accelerated multiplication algorithm by combining integer multiplication and fast reduction. Through implementation, they get good results in reducing chip area and power consumption ($860\mu w$ in 1MHz). They concluded that ECDSA computation requires 511864 clock cycles. Also, \citet{p78} proposed ECDSA processor with RFID tag. This processor has security services (authentication, integrity, and non-repudiation). Authors used this processor to authenticate between tag reader and RFID tag. They used several countermeasures to prevent side channel attacks such as Montgomery ladder algorithm, randomized Projective coordinate and $k$ randomness through SHA-1. Through results, the computation cost of the signature in this processor is 859188 clock cycles (127ms at 6.78 MHz) and the area is 19115 GE. Implementation of ECDSA algorithm is presented with prime field ($F_{p160}$) on RFID \cite{p77}. Authors exchanged SHA-1 hash algorithm with KECCAK hash to reduce running time in ECDSA on RFID. Also, they used a fixed-base comb with w=4 to accelerate and reduce hardware requirements. Results show that their scheme gets an area of 12448 GEs, power consumption 42.42 $\mu w$ in 1MHz and signature generation is less than 140 kcycles. With these results, their scheme competes with other schemes for the binary and prime fields.\\
Recently, the ECDSA with password-based encryption in \cite{p95} has been adopted to improve security and privacy in RFID. Authors pointed out that lightweight processes performed by ECDSA in data signing are significantly effective in RFID. Also, \citet{p96} used ECDSA with the Shamir scheme to secure RFID technology. They applied the Shamir scheme to reduce the cost of two scalar multiplications to one. Finally, ECDSA has been implemented to secure RFID in IoT applications in \cite{p97} have implemented. Authors proposed a shopping system, during the analysis and evaluation, arguing that ECDSA is suitable to sign users requests in the shopping system.
\subsubsection{Smart Cards} 
A smart card is a novel way for authentication as it contains important information for users. This technology is used by several algorithms to implement authentication mechanisms, such as RSA, DSA and ECDSA. ECDSA algorithm is used in this area largely because of the advantages found in ECDSA.\\
The design and implementation of ECC/ECDSA algorithms have been investigated and they are used in constrained-source devices like smart cards \cite{p14}. The authors used a java card that supports the java language and the environment used is next generation integrated circuit card (NGICC). Results showed that the ECC/ECDSA is better than RSA. The results also pointed to the possibility of using these algorithms in the other wireless devices constrained-source. Moreover, the EC algorithms (ECDSA and ECDH) with pairing were used on a Java card \cite{pp88}. ECC requires little data to move between the card reader and the card, compared with the RSA. Furthermore,\citet{p88} concluded that the ECC/ECDSA algorithm is more efficient in terms of power, storage and speed than RSA on the smart card. On the other hand, \citet{p3} discussed using ECDSA with Austrian e-ID (smart card). They noted that ECDSA uses the same keys many times; therefore, they pointed to improved randomness in ECDSA. In 2017, \citet{p98} discussed the security risks of the ECDSA algorithm when applied to a smart card. They proposed the development of a scalar operation algorithm when applying the Montgomery ladder method in ECDSA. They described that ECDSA offers a security solution for their smart card implementations.
\section{Existing Surveys}
In this section, we will present the existing surveys in ECC/ECDSA. In this survey, we will focus on a study to many of the aspects of ECC/ECDSA algorithm. To start with beginning, we will present these articles and then explain the difference between our research and existing studies. Table~\ref{survey:tab2} lists existing surveys for the ECC/ECDSA algorithm. 
\begin{table*}[!t]
\scriptsize
\centering
\caption{Different between existing surveys and our research}
\begin{tabular}{|p{1.0cm}|p{8cm}|p{1.8cm}|p{1.1cm}|}		
		    \hline
			\textbf{Paper}	       & \textbf{Contribution of existing surveys}                          & \textbf{Aspect}       & \textbf{Year} \\ \hline
            \cite{p72}     & Efficiency and flexibility in hardware implementations  &Efficiency    & 2007\\ 			
            \cite{p29}     &  Compared many different signature schemes                &              & 2008\\ \hline
            \cite{p71}	   & Attacks and countermeasures                               &Security      & 2010\\			
            \cite{p48}	   & Hard problems in public cryptography algorithms           &              & 2011 \\
            \cite{p49,p92} & SCA and fault attacks, and countermeasures                &              & 2012,2013\\ \hline
            \cite{p13}     & Security techniques in WSN                                &Applications  & 2015\\ 
            \cite{p52}     & Attack strategies in Bitcoin and Ethereum                 &              & 2016\\ 
            \cite{p100}    & ECC/ECDSA with some applications                          &              & 2017\\ \hline
             This work     & Classification of efficiency, security methods and applications with updated contributions &Efficiency\newline Security\newline Applications   & \\        
            \hline                    
\end{tabular}
\label{survey:tab2}
\end{table*}
\begin{itemize}[noitemsep,nolistsep]
\item \textbf{Performance and efficiency}\\
First of all, in \cite{p72} performance and flexibility have been investigated in the ECC algorithm with accelerators through hardware implementations. Many points in hardware implementations for ECC were discussed such as selecting curves, group law, PM algorithms, and selection of coordinates. In addition, it was pointed out that the architecture of multiple scalar multiplication in ECDSA's verification should supported because this architecture leads to efficiency in hardware's implementations. Much research has pointed out that using hardware's accelerators leads to high performance, but it sacrifices flexibility, where reduction circuits should be used to retrieve the flexibility feature. Similarly, \citet{p29} compared many different signature schemes (ECDSA, XTR-DSA, and NTRUSing) in terms of energy consumption, memory, keys length and signature, and performance. Through implementation, the authors found that NTRUSing algorithm is the best in term performance and memory. However, NTRUSing algorithm suffer from security weakness against attacks.
\item \textbf{Security and countermeasures}\\
A detailed study in \cite{p71} on attacks and countermeasures in ECC algorithm is presented. The authors divided attacks to passive and active attacks. They explained that the countermeasure for a specific attack may be vulnerable to other attacks, whereas countermeasures should have been selected carefully. Therefore, the authors have made some recommendations in selecting countermeasures. Some surveys have studied public cryptography algorithms in terms of computation of hard problems (integer factorization problem(IFP), discrete logarithm problem (DLP), lattices and error correcting codes) in quantum and classical computers \cite {p48}. The authors described RSA, Rabin, ECC, ECDH, ECDSA, ElGamal, lattices (NTRU) and error-correcting code (McEliece cryptography), as they pointed out that ECC provides a higher security level than other cryptosystems; in addition, it presents advantages such as high speed, less storage, and smaller keys sizes. But they did not discuss the use of ECC/ECDSA in applications and implementations of different technologies. Meanwhile, the authors in \cite{p49,p92} explained physical attacks on ECC algorithms, where they focused on two known physical attacks: side channel analysis (SCA), and fault attacks. They also described many attacks including these two types, as they presented countermeasures against these attacks such as simple power analysis (SPA), differential power analysis (DPA) and fault attack (FA) countermeasures. Also, some recommendations were presented for countermeasures that add randomness, countermeasures selection, and implementation issues. However, none of these papers investigated non-physical attacks on public key signature algorithms such as ECDSA.
\item \textbf{Implementation and applications}\\
A study on security techniques has investigated wireless sensor networks (WSNs) \cite{p13}. It focused on three features in WSN security:  key management, authentication, and secure routing. This study pointed out that ECC algorithm was convenient for constrained-resource devices. In addition, a survey on attack strategies was given in relation to ECC and ECDSA algorithms in Bitcoin and Ethereum \cite{p52}. The author pointed out that different standards for curves (such as ANSI X9.63, IEEE P1363, and safecurves), where this survey focused on safecurves with SECP256k1 through using ECDSA, as this paper referred to safecurves as one of the strongest curves standards. The author suggested many basic points to prevent attacks on ECDSA or ECC. Finally, \citet{p100} compared RSA and ECC/ECDSA, and pointed out that ECC/ECDSA exhibited the highest performance with the same level of security from RSA. They noted that 69\% of websites use ECC/ECDSA, 3\% used RSA and the rest used other algorithms. They also described ECC with some applications (such as vehicular communication, e-health and iris pattern recognition). However, they had a duplication between implementation and application. For example, RFID is a technology that can be used to implement a particular application.
\end{itemize}
In our survey, we study the ECDSA algorithm differently to previous studies. First, we integrate three aspects (efficiency, security, and applications) into one search. Second, systematically, we provide different details (ECDSA aspects) of previous studies. Finally, we provide an updated explanation of all these aspects in ECDSA.  
\section{Efficiency Improvement on ECDSA}
 In this section, we will discuss the improvement of ECDSA's efficiency in many ways such as scalar multiplication, coordinate system, and arithmetic operations. In each subsection, successive improvements to several authors will be explained.
\subsection{Efficiency Improvement of Scalar Multiplication}
This section includes many strategies to improve scalar multiplication; these are efficient in terms of scalar representation, curve operations, and ECDSA arithmetic operations.
\subsubsection{Representation Improvement of the Scalar}  
In this section, we explain the scalar representation methods on scalar multiplication. Implementation of scalar multiplication take a large amount of time \cite{p25,p39} in ECDSA and ECC algorithms. Scalar multiplication takes more than 80\% for running time in operation of key computation in sensor devices \cite{p18,p38}. Scalar multiplication (SM) or point multiplication (PM) is a set of point addition $(P+Q)$ and point doubling $(P+P)$ that generates $kP$ (where $\textit{k}$ is a large integer and $P$ is a point on the curve) through $P+P$...($\textit{k}$-times)\cite{p43,pp55}. Representation of 25 points in traditional double-and-add (D\&A) method presents as the following \cite{p42}.\\ 
{\color{white} AA} 25$P$ = 2(2(2($P$+2$P$)))+$P$ \\
Or\\
{\color{white} AA} 25$P$ = 2(2(2(2$P$)+$P$))+(2(2$P$)+$P$)\\ 
The efficient and fast implementation of ECC algorithm and its derivatives are needed to accelerate SM implementation \cite{p44}. 
As we can see from step 5, signature verification in ECDSA algorithm requires 2 scalar multiplications $(u_1G+u_2Q)$ \cite{p50} that require operation of a complexity computation. SM uses three operations: inversion, squaring and multiplication (where it takes the notation i, s and m respectively) that is considered expensive for an ECC algorithm \cite{p43}. These operations are used to evaluate SM efficiently through implementations. When using the Affine coordinate system with SM, both point doubling and point addition consume 2 multiplications, 1 squaring and 1 division operation in field \cite{p43}. Improving on SM leads to reduce computation cost, and running time and thus improve the efficiency of performance in ECC algorithm and its derivatives. The traditional method (double-and-add algorithm) in scalar multiplication uses base 2 in point doubling such as 2(...(2(2$P$))) in addition to point addition \cite{p43}. In this method, point doubling is implemented in each bit in $k$ (where $k$ is integer scalar and represented in binary form) while point addition is implemented when bit equal "1" in $k$ \cite{p50}. Improving SM makes these algorithms convenient for constrained-source devices such as WSN, RFID, and smart card \cite{p23} through reducing of running time and energy consumption. Many researchers have presented methods to represent scalar in order to reduce computation complexity in $kP$. Many representations have used for SM such as traditional method (double-and-add), double base chain (2,3), multibase representation ((2,3,5) and (2,3,7)) and point halving (1/2$P$). All these representations lead to shorter representation length of terms and hamming weight. For instance, multibase representation (2,3,5) is shorter terms length and more redundant than a double base chain (2,3). Representing 160-bit in multibase representation (MBNR) needs 15 terms while double base chain (DBNR) needs 23 terms \cite{p42}. Sometimes, improving on methods of SM representation may bring storage problems, for example, using point halving instead of point doubling with polynomial base requires greater storage in memory \cite{p51}. In the following subsections, we will discuss methods to improve the representation of scalar in SM.
\paragraph{Double-Base Chain (2,3)}
One of the methods used to develop doubling $(2P)$ is tripling $(3P)$. This method was proposed to use bases 2 and 3 to reduce the execution time of the PM. \citet{p53} proposed a point tripling operation $(3P)$ and mixed it with point doubling $(2P)$ depending on various methods to evaluate $2P+Q$ when 1 inversion is more than 6 multiplications that lead to improving tripling. They presented a comprehensive evaluation through i, s and m for operations types: $P+Q$, $2P$, $2P+Q$, $3P$, $3P+Q$, $4P$, $4P+Q$. The authors noted that $2P+Q$ is faster than $P+(P+Q)$ but $2P+Q$ has slightly more cost. They used the idea of Eisentr{\"a}ger  et al. \cite{p67} in removing $y3$ from equations when computing $2P+Q$; the authors used this idea with $3P+Q$ and removed $y4$ when 1i is more than 6m to reduce computation cost. Their results proved that this scheme improved SM efficiency in ECC, ECDSA and ECDH. Double-base chain equation is 
\begin{equation}
k=\sum_{i} s_i2^{b_i}3^{c_i}
\end{equation}
Where $s_i$ is $\pm 1$, and ($2^{b_i}3^{c_i}$) are integer numbers and $b_i$ and $c_i$ decreased monotonically ($b_1 \geqslant b_2...  b_m\geqslant  0$ and $c_1\geqslant  c_2...  c_m\geqslant  0$).
Furthermore, characteristic 3 was investigated with both Weierstrass and Hessian forms \cite{p63}. The authors pointed out that characteristic 3 is efficient in Weierstrass form, where tripling operation performs more efficiently than double and add, while characteristic 3 is not efficient with Hessian form (Triple-and-add (T\&A) used with Weierstrass and double-and-add(D\&A) used with Hessian). In addition, the scalar in the double base number system algorithm (DBNS) is analysed on superlinear EC in characteristic 3 \cite{p66}. Double-base chain is DBNS but with restriction doubling and tripling and increasing the number of point addition through a Horner-like manner. The authors proposed sublinear SM algorithm for PM (i.e. sublinear in scalar's length) with running time O($\frac{log\  n}{log\ log\ n}$), and it is faster than D\&A and T \& A. They pointed out that their algorithm is appropriate to use with large parameters in EC, where selecting large parameters leads to improving performance and security. Also, \citet {p54} proposed double-base chains method through prime and binary fields, using point tripling with a Jacobian coordinate in prime fields, and quadrupling combined with quadruple-and-add with an Affine coordinate in binary fields. They used a modified greedy algorithm to convert $k$ to DBNS form. Also, they applied the same idea that was used in Eisentr{\"a}ger et al. \cite{p67} to evaluate only $x$-coordinate in computation $2P+Q$ but with $4P+Q$ to increase SM efficiency. Through results, the authors have become efficient in speeding SM better than classical D\&A (21\%), NAF (13.5\%) and ternary/binary (5.4\%) in binary fields. In prime fields, their scheme also gets better results than D\&A (25.8\%), NAF (15.8\%), 4NAF (6\%). But results in \cite{p53} are better than this scheme in term inversions (with binary fields) in the computation case 4$P$ and 4$P$+$Q$, where their scheme obtains for 4$P$ (2[i] + 3[s] + 3[m]) and for 4$P$+$Q$ (3[i] + 3[s] + 5[m]). Results in \cite{p53} are 4$P$(1[i] + 5[s] + 8[m]) and 4$P$+$Q$(2[i] + 6[s] + 10[m]) where squaring is free and ignored in binary fields, while their scheme is better than \cite{p53} when using prime fields.
\paragraph{Multibase Representation (2,3,5)}
Multibase representation is a method to improve SM through using a point quintupling (5$P$). It uses 3 bases (2,3,5) called step multibase representation(SMBR).  Multibase representation algorithms are shorter terms, more redundant, and have more sparseness than DBNS algorithms. For instance, representation 160-bit costs 23 terms in bases representation (2,3) whereas 15 terms in representation bases (2,3,5). SMBR equation is presented as follows:
\begin{equation}
k=\sum_{i} s_i2^{b_i}3^{c_i}5^{d_i}
\end{equation}
Where $s_i$ is $\pm 1$, and ($2^{b_i}3^{c_i}5^{d_i}$) are integer numbers and $b_i$,$c_i$ and $d_i$ decreased monotonically ($b_1 \geqslant b_2...  b_m\geqslant  0,c_1\geqslant  c_2...  c_m\geqslant  0$ and $d_1\geqslant  d_2... d_m\geqslant  0$).
\citet{p55} used SMBR with Affine in binary fields and Jacobian in prime fields similar to the scheme in \cite{p54}; therefore, their scheme is a generalization for \cite{p54} but with 3 bases. They improved on a greedy algorithm (mgreedy) in order to fit SMBR and to gain shorter representation and faster running. Moreover, they recommended using small values for exponents because it does not affect cost. Computation of $5P$ was efficient with the prime field through $2(2P)+P$ and $2P+3P$. They found that $2(2P)+P$ costs 9s+17m (with Affine ) and 14s+20m (with Jacobian) while $2P+3P$ costs 22m+12s (with Affine) and 26m+16s (with Jacobian). Moreover, computation of $5P$ efficiency with the binary field (Affine) was 1i+5s+13m.  Through these results, the authors achieved efficiency in SM with prime and binary fields better than previous algorithms such as D\&A, NAF, 3-NAF,4-NAF, ternary/binary and DB chains whether with precomputation or without precomputation points. Finally, \citet{p131} concluded that the cost of $5P+Q$ is 26m + 13s.
\paragraph{Multibase Representation (2,3,7)}
Multibase representation is a scalar representation method, which depends on base triple bases to accelerate point multiplication. It uses 3 bases (2,3,7) and is called multibase number representation(MBNR). This representation is the development of previous representation methods (DBNR (2,3) and SMBR (2,3,5)). \citet{p42} proposed triple base method (binary, ternary and septenary) with addition and subtraction. The greedy algorithm was used to convert an integer to the triple base (2,3,7), through finding the closest integer to a scalar. Their evaluation referred to this method as having better results than previous methods of scalar representation. They pointed out that septupling (7$P$) costs is less than two formulas ($2(2P)+3p$ and $2(3P)+P$), where $7P$=3i + 18m, $2(3P)+P$=4i + 18m and $2(2P)+3P$= 5i + 20m. Also, squaring cost was neglected, regarded as inexpensive in the fields with characteristic 2. MBNR representation was applied through the following equation:
\begin{equation}
k=\sum_{i} s_i2^{b_i}3^{c_i}7^{d_i}
\end{equation}
Where $s_i$ is $\pm 1$, and ($2^{b_i}3^{c_i}7^{d_i}$) are integer numbers and $b_i$,$c_i$ and $d_i$ decreased monotonically ($b_1 \geqslant b_2...  b_m\geqslant  0,c_1\geqslant  c_2...  c_m\geqslant  0$ and $d_1\geqslant  d_2... d_m\geqslant  0$). MBNR (2,3,7) is better than MBNR (2,3,5) in terms of shorter terms length, more redundant and spare. The next example gives a comparison between using quintaupling and septupling when points number= 895712 as mentioned in \cite{p42}.
\begin{gather*}
(using quintaupling)\\
2^43^75^2 + 2^43^55^1 + 2^43^45^0 + 2^13^45^0 + 2^03^25^0 + 2^03^15^0 \\
 (using septupling) \\
2^93^57^1 + 2^73^37^1 + 2^53^17^1 + 2^53^17^0
\end{gather*}
This method presents efficiency in scalar multiplication better than previous representation methods, where we note from the previous example that septupling is more redundant and has fewer terms than quintupling. Also, \citet{p130} proposed MBNR (2,3,5,7) to represent scalar without precomputation. They investigated the costs in their method when a = -3, the cost is 18m + 11s (prime field). The results of their proposal indicate high performance in FPGA implementations with SM at high speed with storage level and execution time.
\paragraph{Point Halving with DBNR and MBNR Representations}
This section will discuss point halving (PH) first, then bases (1/2,3), (1/2,3,5), and (1/2,3,7) respectively. Point halving method is one of the methods used to reduce the cost of point doubling. It means extract $P$ from $2P$ \cite{p43}. Knudsen \cite{p51} and Schroeppel \cite{pp51} separately proposed point halving in order to accelerate point multiplication. They wanted to reduce the cost of computation complexity in point doubling. Point halving was suggested to use instead of all point doubling, where point halving is faster than point doubling in the case of using Affine coordinate on the curve (minimal two-torsion) \cite{p51}. This method was implemented on the binary field (polynomial and normal). Both polynomial and normal basis perform fast computations, but polynomial basis suffers from storage problem. This scheme neglected squaring operation and evaluation as it depends on inversion and multiplication. To implement point halving on binary field in curve, the following equations were used to get $Q=2P$, where
 $P=(x,y)$ and $Q=(u,v)$:
\begin{equation}
\lambda = x + \frac{y}{\lambda}\\
\end{equation}
\begin{equation}
u = (\lambda)^2 + \lambda + a\\
\end{equation}
\begin{equation}
v = x^2  + u(\lambda + 1)
\end{equation}
Through previous equations, we get point halving; the second equation gives us $\lambda$ value, the third equation gives us x and subsequently uses values of x, $\lambda$ to get y value from the first equation. \citet{p57} presented analysis and comparison between SM methods (point doubling and point halving) and used binary fields ($F_{2^{163}}$ and $F_{2^{233}}$) through reduction polynomials of trinomial and pentanomial (using polynomial basis instead of normal) on standards of NIST's FIPS 186-2. Through analysis, they found that point halving is faster (29\%) than point doubling when $P$ is known in advance. Also, they noted $\tau$-adic (Frobenius Endomorphism) used in \cite{p56} faster than PH. They presented a comparison between double-and-add and halve-and-add (H\&A) over $F_{2^{163}}$ through Affine and Projective coordinate systems. Also, they pointed out that point doubling can use with mixed coordinate while point halving must be used in the Affine coordinate. They explained that signature verification in ECDSA has 2 SM and this operation is expensive when $P$ is known. But performing ECDSA verification with halving is more efficient than doubling in addition to halving being better for constrained-resources in the case assumption that storage is available. \citet{p43} mentioned that point halving is faster than point doubling by 5-24\%. \\
Point halving $(1/2P)$ is combined with DBNR (2,3) in \cite{p50} to reduce computation complexity and to increase operations efficiency of scalar multiplication in ECC/ECDSA. The authors implemented their scheme on Pentium D 3.00 GHz using C++ with MIRACL library V 5.0 that deals with a large integer. Equations of scalar representation in this scheme are:
\begin{equation}
k'=2^q mod p
\end{equation}
Where $2^q$ a large integer and value it close to field size.
\begin{equation}
k=k'/2^q=\frac {\sum_{i=1}^{m}s_i 2^{b_i}3^{c_i}}{2^q} = \sum_{i=1}^{m}s_i (1/2)^{q-b_i}3^{c_i}
\end{equation}
Where $s_i$ is $\pm 1$, and ($2^{b_i}3^{c_i}$) are integer numbers and $b_i$ and $c_i$ decreased monotonically ($b_1 \geqslant b_2...  b_m\geqslant  0$ and $c_1\geqslant  c_2...  c_m\geqslant  0$).
The authors showed a comparison between their scheme results and DBNR with using the binary field (163-bit, 233-bit, and 283-bit) and prime field (192-bit) according to NIST's standards. Their scheme achieved better results than an original double-base chain, their results reduced 1/2 inversion, 1/3 squaring and few number of multiplication and that led to improving DBNR in scalar multiplication.\\
In addition, \citet{p43} presented improving on SMBR method through using point halving with SMBR. The original algorithm used bases (2,3,5) while the modified algorithm used (1/2,3,5). The modified scheme adopted point halving and halve-and-add instead of point doubling and double-and-add.  It applied the following equation:
\begin{equation}
k=k'/2^q=\frac {\sum_{i=1}^{m}s_i 2^{b_i}3^{c_i}5^{d_i}}{2^q} = \sum_{i=1}^{m}s_i (1/2)^{q-b_i}3^{c_i}5^{d_i}
\end{equation}
Where $s_i$ is $\pm 1$, and ($2^{b_i}3^{c_i}5^{d_i}$) are integer numbers and $b_i$,$c_i$ and $d_i$ decreased monotonically ($b_1 \geqslant b_2...  b_m\geqslant  0,c_1\geqslant  c_2...  c_m\geqslant  0$ and $d_1\geqslant  d_2... d_m\geqslant  0$).
The greedy algorithm is used to convert an integer to modified MBNR. The authors referred to MBNR (2,3,5) and modified MBNR(1/2,3,5) as having the same terms but costs in modified MBNR (i=7,s=17,m=77) were less than original MBNR (i=15,s=36,m=80) with $k$ = 314159. Sometimes, the original algorithm has fewer terms than the modified MBNR, but modified algorithm remains less costly in operations(i,s,m). The modified and original algorithm are applied on different curves sizes (163, 233, 283), where each curve is tested with 100 random numbers. Results refer to modified scheme of less computation cost (30\%) rather than the original scheme (2,3,5).\\ 
The authors in \cite{p44} proposed combining point halving with MBNR (2,3,7). They used bases (1/2,3,7) instead of (2,3,7). This scheme is less costly than schemes that have bases (2,3,5) and (2,3,7). Also, this scheme is used with the binary field. The authors referred to MBNR as being convenient for ECC because of its shorter representation length and less hamming weight. MBNR (1/2,3,7) representation uses the following equation:
\begin{equation}
k=k'/2^q=\frac {\sum_{i=1}^{m}s_i 2^{b_i}3^{c_i}7^{d_i}}{2^q} = \sum_{i=1}^{m}s_i (1/2)^{q-b_i}3^{c_i}7^{d_i}
\end{equation}
Where $s_i$ is $\pm 1$, and ($2^{b_i}3^{c_i}7^{d_i}$) are integer numbers and $b_i$,$c_i$ and $d_i$ decreased monotonically ($b_1 \geqslant b_2...  b_m\geqslant  0,c_1\geqslant  c_2...  c_m\geqslant  0$ and $d_1\geqslant  d_2... d_m\geqslant  0$).
This scheme presents improvement in the performance of scalar multiplication through reducing computation complexity. Through these results, the authors pointed out that their scheme reduced inversion to 1/2, and squaring to 1/3 and there were fewer numbers of multiplication when compared with previous schemes. Table~\ref{survey:tab3} shows the costs of arithmetic operations to represent scalar.
\begin{table*}[!ht]
\scriptsize
\centering
\caption{Costs of inversion (i), squaring (s), and multiplication (m) for scalar representation}
\label{survey:tab3}
\begin{tabular}{|l|l|l|l|l|l|l|}
\hline
\rowcolor[HTML]{EFEFEF} 
\cellcolor[HTML]{EFEFEF}                                        & \multicolumn{3}{l|}{\cellcolor[HTML]{EFEFEF}Binary field}                            & \multicolumn{3}{l|}{\cellcolor[HTML]{EFEFEF}Prime Field}                             \\ \cline{2-7} 
\rowcolor[HTML]{EFEFEF} 
\multirow{-2}{*}{\cellcolor[HTML]{EFEFEF}Scalar representation} & i      & s      & m                                                                  & i      & s      & m                                                                  \\ \hline
\rowcolor[HTML]{FFFFFF} 
1/2P                                                            & -      & -      & 1                                                                  & -      & -      & -                                                                  \\ \hline
\rowcolor[HTML]{FFFFFF} 
1/2 P + Q                                                       & 1      & -      & 5                                                                  & -      & -      & -                                                                  \\ \hline
\rowcolor[HTML]{FFFFFF} 
P +Q                                                            & 1      & 1      & 2                                                                  & 1      & 1      & 2                                                                  \\ \hline
\rowcolor[HTML]{FFFFFF} 
2P                                                              & 1      & 1      & 2                                                                  & 1      & 1      & 2                                                                  \\ \hline
\rowcolor[HTML]{FFFFFF} 
2P + Q                                                          & 1      & 2      & 9                                                                  & 1      & 2      & 9                                                                  \\ \hline
\rowcolor[HTML]{FFFFFF} 
3P                                                              & 1      & 4      & 7                                                                  & 1      & 4      & 7                                                                  \\ \hline
\rowcolor[HTML]{FFFFFF} 
3P + Q                                                          & 2      & 3      & 9                                                                  & 2      & 3      & 9                                                                  \\ \hline
\rowcolor[HTML]{FFFFFF} 
4P                                                              & 1      & 5      & 8                                                                  & 1      & 9      & 9                                                                  \\ \hline
\rowcolor[HTML]{FFFFFF} 
4P + Q                                                          & 2      & 6      & 10                                                                 & 2      & 4      & 11                                                                 \\ \hline
\rowcolor[HTML]{FFFFFF} 
5P                                                              & 1      & 5      & 13                                                                 & -      & 10     & 15                                                                 \\ \hline
\rowcolor[HTML]{FFFFFF} 
5P + Q                                                          & -      & -      & -                                                                  & 1      & 13     & 26                                                                 \\ \hline
\rowcolor[HTML]{FFFFFF} 
7P                                                              & 3      & 7      & 18                                                                 & -      & 11     & 18                                                                 \\ \hline
\rowcolor[HTML]{FFFFFF} 
7P + Q                                                          & -      & -      & -                                                                  & 1      & 22     & 28                                                                 \\ \hline
\rowcolor[HTML]{FFFFFF} 
1/2 \& DBNR                                                     & DBNR/2 & DBNR/3 & \begin{tabular}[c]{@{}l@{}}Slightly lower\\ than DBNR\end{tabular} & DBNR/2 & DBNR/3 & \begin{tabular}[c]{@{}l@{}}Slightly lower\\ than DBNR\end{tabular} \\ \hline
\rowcolor[HTML]{FFFFFF} 
1/2 \& SMBR                                                     & SMBR/2 & SMBR/3 & \begin{tabular}[c]{@{}l@{}}Slightly lower\\ than SMBR\end{tabular} & SMBR/2 & SMBR/3 & \begin{tabular}[c]{@{}l@{}}Slightly lower\\ than SMBR\end{tabular} \\ \hline
\rowcolor[HTML]{FFFFFF} 
1/2 \& MBNR                                                     & MBNR/2 & MBNR/3 & \begin{tabular}[c]{@{}l@{}}Slightly lower\\ than MBNR\end{tabular} & MBNR/2 & MBNR/3 & \begin{tabular}[c]{@{}l@{}}Slightly lower\\ than MBNR\end{tabular} \\ \hline
\end{tabular}
\end{table*}   
\subsubsection{Methods of Improving Curve Operations}
In this section, we describe the methods which are used to improve the PM by reducing the addition formula operations (point addition and point doubling). In the second part, we explain that some research efforts have used the different methods for first part methods.
\paragraph{Improvement Via Various PM Methods}
Many approaches can be used to improve PM methods because the PM operation consumes a large amount of time from ECC and ECDSA algorithms. Many researchers have presented different methods to improve the performance of basic PM algorithm (D\&A) such as NAF, Window, Comb, Montgomery, Frobenius. In this section, these algorithms will be elaborated. Table~\ref{survey:tab4} shows improvements on SM methods.
\begin{itemize}[noitemsep,nolistsep]
\item \textbf{Double-and-Add (D\&A) Method}\\
D\&A is a basic algorithm in PM. It has been repeatedly used for two operations: point addition$(P+Q)$ and point doubling $(P+P)$. This algorithm is similar to the square-and-multiply algorithm \cite{p18}, and depends on bits in scalar $k$ after it was represent it in binary form, where if bit in $k$ has zero value, the D\&A algorithm will execute point doubling, while if bit in $k$ has one value, this algorithm will execute the point doubling and point addition in each loop. In this method, the number of point addition is a half of the point doubling. D\&A method is explained in algorithm~\ref{alg:algorithm2}.
\begin{algorithm}[t]
\caption{The D\&A algorithm} 
\label{alg:algorithm2}
\begin{algorithmic}
\BState Input: point $P$ on $E(F_p)$; l-bits scalar $k = (k_{l-1} \textrm{...} k_0)_2$
\BState Output: $Q = k.P$
\BState $Q = P$
\For {i = l - 2 downto 0} 
\State $Q= 2.Q$
\If {$k(i)$= 1}  $Q= Q + P$
\EndIf
\EndFor
\BState return $Q$ 
\end{algorithmic}
\end{algorithm}
\item \textbf{Non Adjacent Form (NAF) Method}\\
NAF is the representation of signed-digit and was introduced to reduce execution time in PM; it outperforms D\&A algorithm.  It does not allow for any two nonzero bits in scalar $k$ to be adjacent \cite{p18,p50} and this leads to reducing hamming weight as PM computation depends on the number of zeros and bits length in a scalar. As a result, this algorithm reduces the number of point addition. The NAF method is faster than the square-and-multiply method \cite{p39}. This method can reduce the number of point addition to one-third. It is represented using the following equation:
\begin{equation}
NAF(k)=\sum_{i=0}^{l-1} k_i2^{i}
\end{equation}
Where $k \in \{0,\pm1\}$, NAF method is explained in algorithm~\ref{alg:algorithm3} \cite{b1}.
\begin{algorithm}[t]
\caption{NAF algorithm:} 
\label{alg:algorithm3}
\begin{algorithmic}
\BState Input: A positive integer $k$
\BState Output: NAF($k$)
\BState $i$=0
\While {$k \leq$ 1} 
\If {$k$ is odd}  
\State $k_i$=2−($k$ mod 4)
\State $k=k -k_i$  
\Else 
\State  $k_i=0$ 
\State  $k=k$/2, $i=i$ +1.
\EndIf
\EndWhile
\BState return ($k_i−1, k-i−2, . . ., k_1, k_0$).
\end{algorithmic}
\end{algorithm}
When NAF is combined with a window through a random point in PM with the Jacobian coordinate system, NAF can achieve an improvement of 33.7 addition and 157.9 doubling (with $k$ size is 160-bit) \cite{p60}. Also, \citet{p69} combined simultaneous multiple point multiplication with left-to-right window-NAF, where they convert form ($kP$) for PM to form $k_1p+k_2\phi(p)$($\phi$ which is Endomorphism operation). This method improved point doubling (79 points) and point addition (38 points) that lead to improving PM to 50\% better than traditional methods with $F_{p160}$. Similarly, \citet{p55} made a comparison between their scheme (Quintupling) and NAF algorithm (in addition to many NAF versions such as NAF-3 and NAF-4). Furthermore, NAF algorithm (right-to-left binary method) without precomputations (to use in constrained devices) was used with the mixed coordinate system to improve PM \cite{p62}. The author compared his algorithm with NAF (left-to-right binary method). Results showed that this scheme presented better performance than NAF (left-to-right binary method). Also, \citet{p44} pointed out that multibase non-adjacent form (mbNAF) provided a speeding up of the execution time in PM with improving scalar performance through multibase.
\item \textbf{Frobenius Map Method}\\
Some researchers have used Frobenius map algorithm instead of point doubling $(2P)$ because Frobenius map performs squaring operations $\tau(x,y)=\tau(x^2,y^2)$ where $\tau$ is Frobenius \cite{p56} (that is replacing point doubling with 2 squaring), and is represented by the following equations:
\begin{equation}
2.P= -\tau^{2}P+ \mu\tau P  
\end{equation}
Where $\mu$ = $\pm1$, and
\begin{equation}
k=\sum_{i=0}^{t-1\tau} s_i\tau^{i}
\end{equation}
Where $s_i$ is 0 or $\pm1$ and $\tau$ is Frobenius endomorphism. Scalar $k$ value is obtained through the division repeated for $s$ by $\tau$ \cite{p23} as in the foregoing equation. Algorithm~\ref{alg:algorithm4} shows combining Frobenius map instead of point doubling with D\&A; this algorithm uses addition, subtraction, and Frobenius.
\begin{algorithm}[t]
\caption{Frobenius with D\&A algorithm:} 
\label{alg:algorithm4}
\begin{algorithmic}
\BState Input: Point $P$, $\tau$-adic expansion of $k$ ($k_{l-1},\textrm{...} k_0$).
\BState Output: $k.P$
\BState $P_1 = P$ and $P_2 = 2P$ 
\For {i in 0 to ( l-1) loop} 
\If {$k(i)$ =1}  
\State $Q = Q + P$
\ElsIf {$k(i)$ = -1} 
\State  $Q = Q - P$
\EndIf
\State $P$ = Frobenius($P$)
\EndFor
\BState End loop
\end{algorithmic}
\end{algorithm}
Frobenius endomorphism ($\tau$-adic) was combined with point halving \cite{p56}. The authors used point halving because it is three times faster than point doubling; therefore, they used $\tau$-and-add instead of double-and-add with Koblitz curves; these curves have useful features in acceleration $kP$. They used these curves with fields sizes ($k$-163 bit and $k$-233 bit) to standards of NIST's FIPS in binary fields. $\tau$-NAF was used to reduce point addition from $n/3$ to $2n/7$. 
Results have proved that $\tau$-and-add based $\tau$-NAF provides speeding 14.28\% better than Frobenius method, and in addition, it does not require of additional memory in the case using of normal basis. Frobenius expansion method for the special hyperelliptic curve (introduced by Koblitz) with GLV(Gallant-Lambert-Vanstone) endomorphism (using fields of large characteristic) is presented to improve the efficiency of scalar multiplication \cite{p37}. Through the results, the authors largely improved point doubling while point addition did not improve. This scheme improved scalar point multiplication from 15.6 to 28.3\% when it was implemented on $F_{p^n}$. Reducing the time and increasing performance in ECDSA algorithm is performed through exchange doubling point with Frobenius scheme in the PM \cite{p23}.  The authors presented ECDSA algorithm with curves of subfields Koblitz for binary fields (key length a 163-bit). They explained that binary fields are convenient for hardware implementation.  This scheme achieved good results in performance over time where key generation took time 0.2 ms, signature generation took 0.8 ms and signature verification took 0.4 ms. In addition, this scheme is suited for constrained-source devices such as smart cards, WSN and RFID. In 2017, \citet{p102} submitted a study of the application of twisted Edwards curve (considered efficient models of ECC/ECDSA) with Frobenius endomorphism to reduce point doubling to 50\% for traditional algorithms. They used the 207-bit key (prime field) with two hardware architectures (ASIC and FPGA). They pointed out that their method offers high performance, less memory, and time requirements compared to traditional Frobenius algorithms. Also, they mentioned that integrating their method with a window of 2 size would save 1/16 of the point addition.
\item \textbf{Window Method (WM)}\\
Window algorithm is intended to improve execution time in D\&A method through specific window size (terms computation in a scalar); if window size equals one, the window algorithm is the same as D\&A algorithm. This algorithm uses precomputation for points in the case of the fixed point multiplication (FPM) and also reduced point additions better than D\&A \cite{p18, p25}, as explained in algorithm \ref{alg:algorithm5}.
\begin{algorithm}[t]
\caption{Window algorithm:} 
\label{alg:algorithm5}
\begin{algorithmic}
\BState Input: point $P$ on $E(F_p$), Window width $w$, $d = [l/w], k = (k_{d-1},...,k_0)_2w$.
\BState Output: $Q = kP$.
\BState $P_0=P$ 
\BState Precompute: for i = 1 to $2^{w-1}$ do: $P_i = P_{i-1} + P$
\BState $Q$=0
\For {i = $d$-1 downto 0}
\State $Q$= $2^wQ$
\State $Q$= $Q$ + $P_{kd}$
\EndFor
\BState return $Q$ 
\end{algorithmic}
\end{algorithm}
\citet{p32} used NAF and window methods to improve performance for PM algorithm. They found that window method is more efficient than NAF.  They used hybrid multiplication to reduce access memory instead of multi-precision multiplication. For a practical example, ECC results on MICAz is signature generation=1.3s and signature verification = 2.8s while on TelosB is signature generation=1.6s and signature verification = 3.3s. The authors showed the possibility of using ECC on WSN. Also, the variable-length sliding window was combined with NAF method to reduce points addition (PM) in ECC and ECDSA algorithms \cite{p39}. Computation complexity in PM depends on bits' length and zeros number in an integer. The authors divided elements in NAF($k$) to two windows (non-zero and zero windows) and also divided non-zero window to six types in sliding window. In addition, they used coordinate system (Jacobian) with point doubling and coordinate system (Jacobian Affine) with point addition. Their results demonstrate that their scheme is better and more efficient than NAF, wNAF and square-multiply schemes in terms of efficient point multiplication and time of public key generation. 
\item \textbf{Comb Method (CM)}\\
This method uses binary matrix $(w,d)$ to compute FPM efficiently, where w is row and d is column. It was introduced by Lim and Lee and this algorithm is case special from multi-exponentiation using Straus' trick \cite {b2}. Comb method uses precomputation to improve PM performance \cite{p18,b1}. as in algorithm \ref{alg:algorithm6}.
\begin{algorithm}[t]
\caption{Comb algorithm:} 
\label{alg:algorithm6}
\begin{algorithmic}
\BState Input: A point $P$, an integer $k$, and a window width $w > 2$
\BState Output: $Q = kP$.\\
\BState Precomputation Stage:
\State Compute $[b (w-1), ..., b1, b0]$ $P$ for all $(b (w-1), . . ., b1, b0)\in Z_2^w$ 
\State Write $K = (K(w-1) || . . . || K^1|| K^0)$ padding with 0 on the left if necessary, where each $K^j$ is a bit-string of length $d$.
\State Let $K_{i}^{j}$ denote the $i$-th bit of $K^i$
\State Define $K_i= [K_{i}^{w-1},... ,K_{i}^{1},K_{i}^{0}]$
\State $Q$ = O\\
\BState Evaluation Stage:
\For {$i$ = $d$-1 downto 0} 
\State $Q = 2Q$,
\State $Q=Q+k_iP$
\EndFor
\BState return $Q$ 
\end{algorithmic}
\end{algorithm}
Comb method was modified in \cite{p58} to improve PM through combining comb method with width-$w$ NAF. This method is essentially designed to reduce the number of point addition. It has presented better results than original comb and NAF with comb. Also, it is used to accelerate multiple PM $(kP +rQ)$ in ECDSA algorithm.  The authors explained that their algorithm is convenient for devices constrained-source (memory) when choice suited parameters for an algorithm such as window size. It improved computation complexity from 33\% to 38\% compared to NAF with CM in devices constrained-sources.
\item \textbf{Montgomery Method}\\
Montgomery algorithm eliminates division operation and uses reduction operation efficiently \cite{p32}. This algorithm was introduced by Montgomery and uses only $x$-coordinate and removes $y$-coordinate, and this leads to increasing PM performance.\\
A method introduced by Eisentr{\"a}ger el al. has improved PM's performance with an Affine coordinate in ECC \cite{p67}. This method eliminates field multiplication in the case of using left-to-right of binary SM. It eliminates y-coordinate in addition, doubling and tripling operations. This scheme led to saving field multiplication and thus achieved an improvement of PM's cost from 3.8\% to 8.5\%. For example, in order to perform the operation of $2P+Q$, authors used form $((P+Q)+P)$; also in the operation of $3P+Q$, they used $(((P+Q)+P)+P)$. This trick led to the improvement in PM operation. For instance, with a number of points (1133044P), the cost of original PM is 23i+41s+23m, while the cost of improved PM is 23i+37s+19m, namely, an improvement in m and s. Ciet and Joye \cite{p53} used Eisentr{\"a}ger's idea but with $3P+Q$ when 1i equal more than 6m to reduce the computation cost of PM. Also, Dimitrov et al.\cite{p54} used Eisentr{\"a}ger's idea but with $4P+Q$ to improve the PM efficiency. Montgomery method is represented in algorithm \ref{alg:algorithm7}.
\begin{algorithm}[t]
\caption{Montgomery ladder algorithm:} 
\label{alg:algorithm7}
\begin{algorithmic}
\BState Input: A point $P$ on $E$ and a positive integer $k = (k_{l−1} . . . k_0)_2$.
\BState Output: The point $kP$.\\
\BState $P_1 = P$ and $P_2 = 2P$ 
\For {$i$ = $l$ - 2 down to 0} 
\If {$k_i$ = 0} 
\State $P_1 = 2P_1$ and $P_2 = P_1 \oplus P_2$ 
\Else 
\State  $P_1 = P_1 \oplus P_2$ and $P_2 = 2P_2$
\EndIf
\EndFor
\BState return $P_1$ 
\end{algorithmic}
\end{algorithm}
\citet{p68} suggested that using Montgomery algorithm depends on the parallel-sequential manner to accelerate PM in ECC with $GF(2^{191})$ on Xilinx VirtexE XCV3200 FPG device. Their scheme improved PM's performance to $56.44\mu s$. They compared their scheme with previous schemes (different H/W devices). Results showed that this scheme presents better performance. Also, they pointed out the balance of their scheme between memory size and time. It was suggested to use the Montgomery algorithm with ECDSA (with keys length 224-bit and 256-bit) that has high costs in both computation and communication \cite{p24}. The authors analysed time complexity, where PM consumes a large amount of time from ECDSA's time. They pointed out that Montgomery offers better performance than other algorithms in constrained environments and mobility. \citet{p38} used Montgomery ladder algorithm to point multiplication in ECC (with randomized Projective coordinate) because it provides security against several attacks and performs all operations with $x$-coordinate, thereby increasing PM performance through implementation of ASIC processor with asymmetric cryptography ECDSA ($F_{p192}$). In 2017, \citet{p103} adopted the Montgomery method with the lightweight elliptic curve (twisted Edwards curve (p159, p191, p223, and p255)) to improve speed and balance between memory and performance (cost of communication, execution time, memory) as well as security requirement. They implemented the ECDSA algorithm on Tmote Sky and MICAz nodes. During the implementation, they noted that their scheme offers memory efficiency which requires 6.7k bytes for a SM process instead of 13.1 kbytes for the traditional Montgomery scheme. They recommended the exact selection of ECDSA's parameters curves and the balance between security and efficiency requirements.
\end{itemize}
In summary, the best algorithm for PM (ECC and ECDSA) it is fixed-base comb with w=4 in random binary curves, while in Koblitz curves it is fixed-base window TNAF ($\tau$-adic NAF) with w=6 in case memory is not constrained \cite{p65}. When memory is constrained, Montgomery is the best with random binary and TNAF is the best with Koblitz. The authors pointed out that Koblitz curves are faster than random binary curves ($F_{2^{163}}, F_{2^{233}}$ and $F_{2^{283}}$) for NIST's standards \cite{p65}. Moreover, \citet{p18} analysed in detail the types of SPM (scalar point multiplication) algorithms (that are D\&A, window, and comb method). Examining the results, they concluded that the CM method is faster than D\&A and WM because the CM uses less doubling and adding points, but this method needs more memory. As for security, the D\&A method is the best with 27\%. D\&A is good at memory requirements, needs more time. Therefore, they concluded that several concepts of SPM in ECC algorithm (or ECDSA or ECDH) are based on the user's application and constrained sources. If execution time is a large, the energy consumption increases. The authors carried out the Secure-CM and earned good time to implement (1.57 s); as demonstrated ECC is applicable in WSN.
\paragraph{Improvement of Efficiency Via Other Methods}
In this subsection, we will explain a set of different ideas from previous methods to improve ECDSA performance through reducing consumption time in signature generation and signature verification.\\
A method was suggested to accelerate signature verification in ECDSA algorithm \cite{p40}. The authors used a number of small bits (side information (1 or 2 bits)) to specify the number of allowed points in point multiplication; through the modified algorithm, points double was largely reduced. They implemented traditional ECDSA algorithm and modified ECDSA algorithm on ARM7TDMI platform processor with a finite field $F_{p384}$ by NIST, and achieved the result that modified ECDSA verification is 40\% better than traditional ECDSA. They discovered that this changing does not affect ECDSA standards and proved that their scheme has the same security as traditional ECDSA. Also, an addition formula was proposed through using Euclidean Addition Chains (EAC) with Fibonacci to avoid difficult to find small chain \cite{p64}. The author has used a Fibonacci-and-add algorithm instead of double-and-add (using Jacobian coordinate and characteristic greater than 3). Some improvements were added to this scheme through the window and signed representation to improve algorithm performance. The author has compared his scheme with many schemes (D\&A, NAF, 4-NAF and Montgomery ladder). Results indicated that his scheme outperforms  D\&A only in the case of improvement addition (window or signed representation). In addition, the improvement of signature verification in ECDSA is achieved through cooperating of adjacent nodes in the computation of intermediate results \cite{p46}. The proposed scheme used 1PM+1Add in nodes that use intermediate results instead of 2PM+1Add as in the original scheme. The authors analysed performance and security in the scheme of signature verification with many attacks (independent and collusive). They noted that these attacks do not have a large effect on their scheme. The modified scheme has saved energy consumption 17.7-34.5\% better than the traditional scheme, as signature verification in the modified scheme is 50\% faster than the original scheme. They implemented their scheme on Micaz motes with the finite field ($F_{2^{163}}$).\\ 
Furthermore, \citet{p25} proposed a scheme to improve scalar multiplication by reducing point additions. They presented a method to generate $k$ depend on the generation of an integer S periodically. They achieved good results in improving point addition $\sqrt{3/4 \times L}$ whereas binary scheme ($L$/2). Point doubling in their scheme is similar to previous schemes (D\&A, NAF, WM). Their scheme does not require additional memory and depends on the growth rate of a small with bit length growth in scalar($k$). Also, it is appropriate to implement in H/W because it needs simple operands. Because scalar multiplication (ECDSA) consumes a large time through processing, that leads to power consumption \cite{p41}. The authors exchanged linear point multiplication method with their method using divide and conquer algorithm. This algorithm uses a binary tree to quotient values and a skew tree to reminder values, where it processes points through parallel computation method. Results showed that they achieved better efficiency than linear scalar multiplication in terms of a number of clock pulses and power consumption. In 2017, \citet{p101} have proposed a method of random scalar that relies on the covering systems of congruence relationships. A random was applied in a scalar representation using mixed-radix SM algorithm. They generated n-covers randomly with a greedy method, depending on available memory as well as recommended selection of a large n. Also, they pointed out that their method is more efficient, less expensive and incurs fewer additional expenses for arithmetic operations than the Coron’s randomization, D\&A, NAF, and wNAF methods. Table~\ref{survey:tab4} shows improvements on other SM methods.
\begin{table*}[!ht]
\scriptsize
\centering
\caption{Improvements on different SM methods}
\label{survey:tab4}
\begin{tabular}{|l|l|l|l|}
\hline
\rowcolor[HTML]{EFEFEF} 
Paper         & SM methods                                                               & Subsequent improvements of SM (PA and PD)                                                                    & Field           \\ \hline
\rowcolor[HTML]{FFFFFF} 
\cite{p39} & Traditional NAF                                                  & 1/3 PA      & $F_p$                                                                                                               \\ \hline
\rowcolor[HTML]{FFFFFF} 
\cite{p60} & wNAF                                                             & \begin{tabular}[c]{@{}l@{}}Random point: PA (33.7) and PD (157.9)\\ \\ Fixed point: PA (30) and PD (15)\end{tabular}  & $F_{p160}$                                                                                                               \\ \hline
\rowcolor[HTML]{FFFFFF} 
\cite{p69} & \begin{tabular}[c]{@{}l@{}}wNAF with\\ Endomorphism\end{tabular} & Roughly 50\% SM   & $F_{p160}$                                                                                                                                                                                                                    \\ \hline
\rowcolor[HTML]{FFFFFF} 
\cite{p44} & mbNAF                                                            & 50\% SM     & $F_p$ and $F_{2^m}$                                                                                                                                                                                                                                                                                                                                \\ \hline \hline
\rowcolor[HTML]{FFFFFF} 
\cite{p18}  & Traditional D\&A                                                                & PA is 1/2 PD                                                                            & $F_p$              \\ \hline
\rowcolor[HTML]{FFFFFF} 
\cite{p56}  & \begin{tabular}[c]{@{}l@{}}$\tau$-NAF\\ \\ $\tau$-NAF with PH\end{tabular}      & \begin{tabular}[c]{@{}l@{}}1/3 PA and reduce PD\\ \\ 2n/7 PA and reduce PD\end{tabular} & $F_{2^m}$  (163,233) \\ \hline
\rowcolor[HTML]{FFFFFF} 
\cite{p37}  & Frobenius with GLV                                                              & 28.3\% SM                                                                               & $F_p$              \\ \hline
\rowcolor[HTML]{FFFFFF} 
\cite{p102} & \begin{tabular}[c]{@{}l@{}}Frobenius with \\ twisted Edwards curve\end{tabular} & 3n/4 PA and 1/2 PD                                                                      & $F_{p207}$          \\ \hline \hline
\rowcolor[HTML]{FFFFFF} 
\cite{p25} \newline & Traditional window                                                                & Reduce PA and PD compared with NAF & $F_p$ (192,256,512)             \\ \hline
\rowcolor[HTML]{FFFFFF} 
\cite{p32}  &   \begin{tabular}[c]{@{}l@{}} Sliding window \\ \end{tabular}  & 10\% better than NAF & $F_{p512}$             \\ \hline
\rowcolor[HTML]{FFFFFF}
\cite{p39}  & \begin{tabular}[c]{@{}l@{}}Variable-length \\ sliding window\end{tabular}     & 27.4\% better than wNAF & $F_p$             \\ \hline \hline
\rowcolor[HTML]{FFFFFF} 
\cite{p18} & Traditional cm                                                        & \begin{tabular}[c]{@{}l@{}}cm is less PA and PD\\ than wm and D\&A\end{tabular} & $F_p$      \\ \hline
\rowcolor[HTML]{FFFFFF} 
\cite{p58} & \begin{tabular}[c]{@{}l@{}}cm with NAF\\ \\ cm with wNAF\end{tabular} & \begin{tabular}[c]{@{}l@{}}33\% SM\\ \\ 38\% SM\end{tabular}                    & $F_{p160}$ \\ \hline \hline
\rowcolor[HTML]{FFFFFF} 
\cite{p67}  & Traditional Montgomery                                                          & \begin{tabular}[c]{@{}l@{}}8.5\% SM\\ memory size (SM)=13.1k\end{tabular}                                                                                                                                                                 & $F_p$ (160,256)         \\ \hline
\rowcolor[HTML]{FFFFFF} 
\cite{p54}  & Improved Montgomery                                                             & \begin{tabular}[c]{@{}l@{}}Prime field: 21\% SM is better than Q \& A\\                     13.5\% SM is better than NAF\\ Binary field: 25.8\% SM is better than Q \& A\\                      15.8\% SM is better than NAF\end{tabular} & $F_p$ and $F_{2^{m}}$     \\ \hline
\rowcolor[HTML]{FFFFFF} 
\cite{p53}  & Improved Montgomery                                                             & \begin{tabular}[c]{@{}l@{}}Reduce PA and PD in \\ traditional Montgomery\end{tabular}                                                                                                                                                     & $F_p$                   \\ \hline
\rowcolor[HTML]{FFFFFF} 
\cite{p68}  & \begin{tabular}[c]{@{}l@{}}Montgomery with\\ parallel sequential\end{tabular}   & roughly 51\% SM with time=0.056ms                                                                                                                                                                                                         & $F_{2^{191}}$             \\ \hline
\rowcolor[HTML]{FFFFFF} 
\cite{p103} & \begin{tabular}[c]{@{}l@{}}Montgomery with\\ twisted Edwards curve\end{tabular} & \begin{tabular}[c]{@{}l@{}}Faster SM than traditional Montgomery\\ memory size (SM)=6.7k\end{tabular}                                                                                                                                     & $F_p$ (159,191,223,255) \\ \hline \hline
\rowcolor[HTML]{FFFFFF} 
\cite{p40}  & SM with side information                                                 & 40\% is better than traditional verification                                                                 & $F_{p384}$      \\ \hline
\rowcolor[HTML]{FFFFFF} 
\cite{p64}  & SM with Fibonacci                                                        & \begin{tabular}[c]{@{}l@{}}Improve PA in D\&A and reduce\\ PA cost (10\%)\end{tabular}                       & $F_{2^m}$       \\ \hline
\rowcolor[HTML]{FFFFFF} 
\cite{p46}  & \begin{tabular}[c]{@{}l@{}}SM with intermediate\\  results\end{tabular}  & Improve signature verification (50\%)                                                                        & $F_{2^{163}}$   \\ \hline
\rowcolor[HTML]{FFFFFF} 
\cite{p25}  & \begin{tabular}[c]{@{}l@{}}SM with integer S\\ periodically\end{tabular} & \begin{tabular}[c]{@{}l@{}}Improve PA ($\sqrt{3/4 \times L}$ is better\\ than D\&A, NAF, and WM\end{tabular} & $F_p$ (192,256) \\ \hline
\rowcolor[HTML]{FFFFFF} 
\cite{p41}  & SM with binary tree                                                      & Reduce SM time and complex computations                                                                      & $F_{2^m}$       \\ \hline
\rowcolor[HTML]{FFFFFF} 
\cite{p101} & SM with random n-cover                                                   & \begin{tabular}[c]{@{}l@{}}Reduce SM cost is better than D\&A, NAF, \\ and wNAF\end{tabular}                 & $F_p$           \\ \hline 
\end{tabular}
\end{table*}
\subsection{Efficiency Improvement of Coordinate Systems}
ECDSA has to perform complex operations; these operations consume resources in constrained devices. Using the appropriate coordinate system, could lead to reducing costs of high computation in doubling and addition operations. The improvement of computation complexity in PM depends on point representation that is considered important in curve operations \cite{p79}. Many different coordinate systems used with these algorithms such as Affine, Projective, Jacobian, Chudnovsky Jacobian, modified Jacobian, and mixed. Coordinate systems are used to represent and contribute to speeding computation in ECC/ECDSA algorithm. There is no coordinate system to accelerate both point addition and doubling \cite{p61}. Some coordinate systems require faster or slower computation time in the point doubling and point addition operations than other coordinate systems; for instance, point doubling uses less computation in Jacobian while the point addition uses more computation in the Jacobian than Projective and Chudnovsky Jacobian coordinates.
\begin{itemize}[noitemsep,nolistsep]
\item \textbf{Using Affine Coordinate}\\
Affine is a basic coordinate system that is used with doubling and addition operations. This system uses two coordinates $(x,y)$. When point addition $(P+Q)$ uses the Affine system, then the cost of the resulting point ($R$) is 1 inversion, 2 multiplication and 1 squaring, while cost($R$) in the case point doubling is 1 inversion, 2 multiplication, and 2 squaring \cite{p59}. Costly arithmetical operations in point addition and point doubling are multiplication, squaring and inversion. Inversion operation is much lower than multiplication operation. The Affine coordinate system uses inversion operations \cite{p32}. Inversion operation is required in point addition and point doubling when using Affine coordinate. But Affine coordinate needs fewer multiplication operations than other coordinate systems such as Projective coordinate \cite{p60}.  
\item \textbf{Using Projective Coordinate}\\
Inversion operation is dramatically expensive in point addition and point doubling. The Projective coordinate system removes this operation and that leads to improving PM performance in ECC/ECDSA. This system uses three coordinates $(X,Y,Z)$ and $Z\neq$0 \cite{p59}, which take formula $(X/Z,Y/Z)$ corresponding to Affine coordinate. The Projective coordinate is represented as the following equation (with prime fields) \cite{p18}:
\begin{equation}
Y^2Z=X^3+aXZ^2+bZ^3
\end{equation}
L{\'o}pez and Dahab also used three coordinates but using formula $(X/Z,Y/Z^2)$ and $Z\neq$0 corresponding to Affine coordinate \cite{pp18}. Their scheme improved original Projective on a curve in the binary field. Furthermore, Projective coordinate is used with random binary and koblitz curves ($F_{2^{163}},F_{2^{233}}$ and $F_{2^{283}}$) for NIST's standards instead of Affine \cite {p65}. This coordinate leads to significant improvement in PM ECC and ECDSA. Then, \citet{p38} used randomized Projective coordinate through generating a randomized number on x-coordinate in base point. Also, different coordinate systems have been discussed, such as Projective and Affine coordinates \cite{p18}. Through experiments, the authors found that Projective coordinates are faster by 91\% than Affine coordinates but Projective coordinate avoids inversion operation, and vice versa for the memory, while the Projective coordinates require more memory because they use three coordinates $(X,Y,Z)$. \citet{p79} proposed $\lambda$-Projective coordinate through using three coordinates $(X,Y,Z)$ with binary field $(F_{2^{254}})$ to improve computation efficiency in PM. The formula used for coordinates is $(X/Z,L/Z)$ corresponding with $\lambda$-Affine formula, where the $\lambda$-Affine formula is $(X,X+Y/X)$. $\lambda$-Projective improved L{\'o}pez and Dahab Projective(LD-Projective), where the $P+Q$ cost in $\lambda$-Projective is 11m+2s while in LD-Projective it is 13m+4s. Also, $2P+Q$ cost in $\lambda$-Projective is 10m+6s while in LD-Projective is 11m+10s. In addition, $\lambda$-Projective improved H\&A to 60\% in squaring and 6\% in multiplication better than LD-Projective. Their scheme was implemented on Sandy Bridge platform. The authors achieved computation 69500 clock cycles through the single core with H\&A and 47900 clock cycles through the multi core combining Gallant, Lambert, and Vanstone, (GLV) technique, H\&A and D\&A in SM unprotected. Also, they achieved computation 114800 clock cycles through single core protected. Their scheme achieves performance better by 2\% than Ivy Bridge and 46\% than Haswell platforms. In 2017, \citet{p104} proposed a parallel SM method based on L{\'o}pez-Dahab Projective that required 4 multiplications for DBL and 16 multiplications for ADD. His scheme applied on FPGA and AT$^2$ processors and obtained high speed for SM's performance when this scheme was implemented on eight processors with L{\'o}pez-Dahab Projective.
\item \textbf{Using Jacobian Coordinate}\\
This coordinate system does not require inversion in addition formula (addition and doubling), but uses inversion only in the final computation stage. It is similar to Pojective coordinate (also it is an improvement of Projective coordinate). Jacobian provides less running time in point doubling and more running time in point addition than Projective coordinate \cite{p60}. This system uses three coordinates $(X,Y,Z)$ and $Z\neq$0, which take formula $(X/Z^2,Y/Z^3)$ corresponding for Affine coordinate \cite{p59}.
\begin{equation}
Y^2Z=X^3+aXZ^4+bZ^6
\end{equation}
Due to the point addition, Jacobian consumes more computation time than other coordinate systems. Chudnovsky Jacobian coordinate is proposed to accelerate point addition in the Jacobian coordinate. This system uses coordinates $(X,Y,Z,Z^2,Z^3)$ and saves 1m+1s in point addition \cite{pp62}. This coordinate has a disadvantage that point doubling is slower than Jacobian coordinate \cite{p62}. The author pointed out that Chudnovsky is the fastest in point addition and modified Jacobian is the fastest in point doubling. A Jacobian coordinate is used to improve running time in EC exponentiation with fixed point and random point. This coordinate is better than Affine and Projective coordinates \cite{p60}. Also, modified Jacobian coordinate system was able to achieve better performance than Affine, Projective, Jacobian and Chudnovsky Jacobian coordinate systems, where Affine has addition cost (i+2m+s) and doubling cost(i+2m+2s), Projective has addition cost (12m+2s) and doubling cost (7m+5s), Jacobian has addition cost (12m+4s) and doubling cost (4m+6s), Chudnovsky Jacobian has addition cost(11m+3s) and doubling cost (5m+6s) and, modified Jacobian has addition cost (13m+6s) and doubling cost (4m+4s) \cite {p61}. This coordinate system is applied on EC exponentiation through prime field ($F_{p160}, F_{p192}$ and $F_{p224}$). Modified Jacobian is faster point doubling than all previous coordinate systems. In addition, the modified Jacobian coordinate is used to represent addition formula when ECDSA algorithm is implemented as authentication protocol in wireless on processor ARM7TDMI \cite{p70}. The authors implemented ECDSA with $F_p$ with different field sizes ($F_{p160}$, $F_{p176}$, $F_{p192}$, $F_{p208}$ and $F_{p256}$) depending on standards of ANSI X9F1 and IEEE P1363. Through results, they get running time 46.4ms for signature generation and 92.4ms for signature verification when using $F_{p160}$. They pointed out that their scheme improved bandwidth and storage compared with previous schemes. \citet{p59} pointed out that Jacobian is faster than Affine, Projective, and Chudnovsky in point doubling. Through results, these coordinate systems improved speeding of computation in ECC/ECDSA through reducing execution cycle.
\item \textbf{Using Mixed Coordinate Systems}\\
The mixed coordinate system uses more one coordinate system to represent addition formula (which each point uses a different system) \cite{p61,p62} to get the best performance and least computation time in PM. Many different mixed coordinate systems such as Jacobian-Affine, Affine-Projective, and Chudonvsky-Affine \cite{p59}.\\
\citet{p61} pointed out that modified Jacobian mixed with other coordinate systems (Jacobian and Affine or Jacobian and Chudnovsky Jacobian) presents a significant improvement on computation time with the prime field ($F_{p160}, F_{p192}$ and $F_{p224}$). Execution time is reduced using Jacobian-Chudnovsky coordinate with prime field (when using $w$=5 for $F_{p192}$, $F_{p224}$ and $F_{p256}$, and $w$=6 for $F_{p384}$ and $F_{p521}$). Also, Jacobian and Chudnovsky coordinates improve PM performance and Chudnosky is preferred although that requires some extra storage (precomputation for points) \cite {p59}. Different coordinate systems were investigated (Affine, Projective, Jacobian and LD Projective) on point addition and doubling with characteristic 3 in both Weierstrass and Hessian forms \cite{p63}. The authors noted that Jacobian is the most efficient with Weierstrass while Projective is the most efficient with Hessian. Also, they used mixed coordinate (Affine-Projective) on both Weierstrass and Hessian. Through using mixed coordinate, results indicate improved timing in addition formula. Mixed coordinate system (Affine-Jacobian) is used to reduce m and s operations or to avoid i operations \cite{p32}. Authors pointed out that mixed coordinate is better performance (6\%) than Jacobian coordinate. In the mixed coordinate, point addition consumes 8m and 3s (12 modular reductions) while point doubling consumes 4m and 4s (11 modular reductions). Also, mixed coordinate is used with right-to-left NAF algorithm \cite{p62}, where applied Jacobian is used with point addition and modified Jacobian with point doubling. The author pointed out that modified Jacobian is fast in point doubling but is slow in point addition; therefore, the author used Jacobian in point addition. Through results, this scheme presented efficiency of a scalar by 13.33$\ell$m (where $\ell$ is bits' length in a scalar) compared with Jacobian (15.33$\ell$m) and modified Jacobian (14.33$\ell$m). But when three fields (temporary) variables additional are used, then the left-to-right binary method described in \cite {p61} was a better performance. Jacobian-Affine is faster than other coordinate systems in point addition. Using Jacobian coordinate in point doubling and Jacobian-Affine in point addition \cite{p39}. Authors pointed out that Jacobian is the fastest in point doubling and mixed coordinate is the fastest in point addition. \citet{p79} used mixed coordinate to compute $2P+Q$ where $\lambda$-Affine was used to represent $P$ point and $\lambda$-Projective to represent $Q$ point. They achieved computation cost less than LD-Projective. In 2017, \citet{p105} indicated that mixed Jacobian-Affine coordinate is more efficient than Jacobian coordinate. They applied a mixed coordinate with ECDSA to improve the efficiency of computation processes to authenticate users during online registration. Their results indicated that mixed coordinate reduces calculations in both point addition and point doubling. Table~\ref{survey:tab5} shows the successive improvements of different coordinate systems in SM such as Affine, Projective, Jacobian, and mixed coordinates.
\end{itemize}
\begin{table*}[!ht]
\scriptsize
\centering
\caption{Improvements of different coordinate systems}
\label{survey:tab5}
\begin{tabular}{|l|l|l|l|l|}
\hline
\rowcolor[HTML]{EFEFEF} 
Paper             & Coordinate system                                                                    & \begin{tabular}[c]{@{}l@{}}Subsequent improvements of \\ SM (PA and PD)\end{tabular}                                                                               & Coordinate formula                                                                                & Field                                                                     \\ \hline
\rowcolor[HTML]{FFFFFF} 
\cite{p60}      & Affine                                                                               & \begin{tabular}[c]{@{}l@{}}1i+2m (PA) and 1i+2m (PD)\\ m is less than Projective\end{tabular}                                                                      & ($X$,$Y$)                                                                                             & $F_p$                                                                     \\ \hline \hline
\rowcolor[HTML]{FFFFFF} 
\cite{p59,p18}  & Traditional Projective                                                               & \begin{tabular}[c]{@{}l@{}}12m (PA) and 7m (PD)\\  i is removed\\ 91\% is faster than Affine\end{tabular}                                                          & ($X/Z$,$Y/Z$)                                                                                         & \begin{tabular}[c]{@{}l@{}}$F_p$\\  (192,224,256,384,521)\end{tabular}    \\ \hline
\rowcolor[HTML]{FFFFFF} 
\cite{pp18}     & LD-Projective                                                                        & \begin{tabular}[c]{@{}l@{}}17\% (SM) is better than\\  traditional Projective\end{tabular}                                                                         & ($X/Z$,$Y/Z^2$)                                                                       & $F_{2^m}$                                                                 \\ \hline
\rowcolor[HTML]{FFFFFF} 
\cite{p65}      & \begin{tabular}[c]{@{}l@{}}Traditional Projective\\ \\ LD-Projective\end{tabular}    & \begin{tabular}[c]{@{}l@{}}13m (PA) and 7m (PD)\\ \\ 14m (PA and 4m (PD)\end{tabular}                                                                              & \begin{tabular}[c]{@{}l@{}}(X/Z,Y/Z)\\ \\ (X/Z,Y/Z\textasciicircum 2)\end{tabular}                & \begin{tabular}[c]{@{}l@{}}$F_{2^m}$ \\ (163,233,283)\end{tabular}        \\ \hline
\rowcolor[HTML]{FFFFFF} 
\cite{p79}      & $\lambda$-Projective                                                                 & \begin{tabular}[c]{@{}l@{}}LD-Projective:                        \\       13m (PA) and 11m (PD)\\ $\lambda$-Projective:\\       11m (PA) and 10m (PD)\end{tabular} & ($X/Z$,$L/Z$)                                                                                         & $F_{2^{254}}$                                                             \\ \hline
\rowcolor[HTML]{FFFFFF} 
\cite{p104}     & \begin{tabular}[c]{@{}l@{}}LD-Projective with\\ parallel SM\end{tabular}             & Improved SM execution time                                                                                                                                         & ($X/Z$,$Y/Z^2$)                                                                       & $F_{2^{163}}$                                                             \\ \hline \hline
\rowcolor[HTML]{FFFFFF} 
\cite{p59,p18}  & Traditional Jacobian                                                                 & 12m (PA) and 4m (PD)                                                                                                                                               & ($X/Z^2$,$Y/Z^3$)                                                     & \begin{tabular}[c]{@{}l@{}}$F_p$ \\ (192,224,256,384,521)\end{tabular}    \\ \hline
\rowcolor[HTML]{FFFFFF} 
\cite{p62}      & Chudnovsky Jacobian                                                                  & 11m (PA) and 5m (PD)                                                                                                                                               & ($X,Y,Z,Z^2$,$Z^3$)                                                   & $F_{2^m}$                                                                 \\ \hline
\rowcolor[HTML]{FFFFFF} 
\cite{p61,p70}  & Modified Jacobian                                                                    & 13m (PA) and 4m (PD)                                                                                                                                               & ($X$,$Y$,$Z$,$aZ^4$)                                                                      & $F_p$ (160,192,224)                                                       \\ \hline \hline
\rowcolor[HTML]{FFFFFF} 
\cite{p63}      & Affine-Projective                                                                    & \begin{tabular}[c]{@{}l@{}}Weierstrass form:\\        9m (PA) and 1i+2m (PD)\\ Hessian form:\\        10 (PA) and 1i+5m (PD)\end{tabular}                          & \begin{tabular}[c]{@{}l@{}}($X$,$Y$) and\\ ($X/Z$,$Y/Z$)\end{tabular}                                     & \begin{tabular}[c]{@{}l@{}}$F_{3^{97}}$ and \\ $F_{2^{163}}$\end{tabular} \\ \hline
\rowcolor[HTML]{FFFFFF} 
\cite{p32}      & Affine-Jacobian                                                                      & \begin{tabular}[c]{@{}l@{}}8m (PA) and 4m (PD)\\ 6\% is better performance\\  than Jacobian\end{tabular}                                                           & \begin{tabular}[c]{@{}l@{}}($X$,$Y$) and\\ ($X/Z^2$,$Y/Z^3$)\end{tabular} & $F_{p160}$                                                                \\ \hline
\rowcolor[HTML]{FFFFFF} 
\cite{p62}      & \begin{tabular}[c]{@{}l@{}}Jacobian- modified \\ Jacobiann\end{tabular}              & 11m (PA) and 3m (PD)                                                                                                                                               & ($X/Z^2, X/Z^3$)                                                     & $F_{2^m}$                                                                 \\ \hline
\rowcolor[HTML]{FFFFFF} 
\cite{p39,p105} & \begin{tabular}[c]{@{}l@{}}Jacobian-Affine and \\ Jacobian\end{tabular}              & 8m (PA) and 4m (PD)                                                                                                                                                & \begin{tabular}[c]{@{}l@{}}($X$,$Y$) and\\ ($X/Z^2,Y/Z^3$)\end{tabular} & $F_p$ (224,256)                                                           \\ \hline
\rowcolor[HTML]{FFFFFF} 
\cite{p79}      & \begin{tabular}[c]{@{}l@{}}$\lambda$-Projective \\ and $\lambda$-Affine\end{tabular} & 8m (PA) and 4m (PD)                                                                                                                                                & \begin{tabular}[c]{@{}l@{}}($X/Z$,$L/Z$) and\\ ($X$,$X+Y/X$)\end{tabular}                                 & $F_{2^{256}}$                                                             \\ \hline
\end{tabular}
\end{table*}
\subsection{Efficiency Improvement Via Algorithm Simplification}
This section will show some different approaches to improving ECDSA performance. A set of ideas which researchers used to simplify the algorithm to improve the performance ECDSA includes removing the inversion in the generation of the signature and the signature verification, a collection of signatures, using a few keys and using two points instead of shared three curved points.\\ ECDSA algorithm is mainly used to provide integration, authorisation, and non-repudiation depending on servers' CA (certification authority) \cite{p21}. The authors added an ID, ECDSA threshold and trust value with the public key to authorise node to gain access to the information. Also, it was stated that data transmission of ECDSA in the wireless sensor network (WSN) consumes more energy than computation operations \cite{p15}. To reduce energy consumption, in data transmission, all encrypted data aggregated from all sensor nodes are directly sent to the Base station without decryption. This operation leads to reducing energy consumption and increasing network lifetime. Encryption and signing algorithms are used for the collected data \cite{p11,p15}. Encryption algorithm EC-OU (Elliptic Curve Okamoto-Uchiyama) is used to maintain confidentiality and ECDSA algorithm to maintain integration, both algorithms are used during homomorphic. The idea behind this scheme is that encryptions, signatures and public keys gathered together from all nodes are sent to aggregator (CH), then these aggregators are forwarded to the top level (parent aggregators) and so on up BS, that is, it is not the decryption or signature verification in the aggregators, but these processes are carried only in BS, that has a high capacity. The authors concluded that this scheme offers energy and time efficiency due to a collection of encryptions and signatures that are sent to the BS. Also, this scheme avoids the signature verification process that takes more time than the signature generation. In addition, it is efficient for large networks. Also, ECDSA algorithm was investigated in detail \cite{p12}, and contains some of the problems that were related to using inversion processes in generation and verification of the signature. An inversion process consumes a great deal of time, which affects the calculation. This process consumes 10 times the multiplication process. In this scheme, the inversion process was removed, and the results demonstrated that this scheme is more efficient than the original ECDSA scheme. It had less time and therefore less computation and energy. This scheme was applied to the sensor node Micaz.\\ 
Using a few keys saved energy, memory and reduced the computation in ECDSA algorithm \cite{p8}. Researchers used ECDSA to secure the connection between the gateway and the cluster head(CH), and CH and nodes. This scheme broadcasts only session keys ($km$), which reduces energy, as the session keys are deleted after their account. This operation leads to less memory (only a few keys stored in memory). The public key of the gateway does not require power, and calculation processes are accomplished through the use of a random number and hash function that require less computation operations. This scheme used periodic authentication in session key. Therefore, it prevents attacks and is appropriate for environments of constrained-source. Finally, modified ECDSA algorithm is proposed through using two curve points publicly (public key $Q$ and point of signature verification $(x, y)$) \cite{p75}, where the base point $G$ becomes the private point. In this algorithm, $s$ parameter only is used (without using $r$ parameter) thus removing $r$ overheads. The authors pointed out that the modified ECDSA algorithm is less complex (more performance) using less number of PMs, point addition and point doubling than original ECDSA algorithm.
\titlespacing*{\section}
{0pt}{6.0pt plus 0pt minus 0pt}{6pt plus 0pt}
\titlespacing*{\subsection}
{12pt}{6pt plus 6pt  minus 6pt}{3pt plus 6pt  minus 6pt}
\titlespacing*{\subsubsection}
{12pt}{6pt plus 6pt  minus 6pt}{3pt plus 6pt  minus 6pt}
\titlespacing*{\paragraph}
{12pt}{6pt plus 6pt  minus 6pt}{3pt plus 6pt  minus 6pt}
 \section{Security Improvement on ECDSA}
In this section, we will consider the security in the ECDSA algorithm. A list of mechanisms for improving the security in ECDSA is given. In the beginning, we will come across types of attacks, security requirements and countermeasures by categorizing them into non-physical and physical attacks, as each category includes passive and active attacks. We then will offer security mechanisms to enhance and protect ECDSA's signatures from tampering.
\subsection{Categories of Attacks}
First of all, we give a review of the attacks that threaten the security level in the ECDSA algorithm. Attacks are becoming increasingly sophisticated, so at the same time there should be countermeasures against such attacks. But these countermeasures are expensive in terms of time, storage, and complex computations. Also, it is difficult to develop a countermeasure for each attack \cite{p73}.\\
When the ECDSA algorithm was implemented, it needed to use countermeasures against known attacks \cite{p27}.
The use of standard criteria to credible organizations such as IEEE, ISO, NIST, NSA, FIPS, and ANSI is extremely important to prevent many attacks. Namely, the abnormal curves are weak for attacks. In addition, the parameters' validation of ECDSA leads to strong security against attacks \cite{p22}. Protection of the private key $(d)$ and ephemeral key $(k)$ in the ECC/ECDSA algorithm are essential procedures because if an adversary could get these the keys, the attacker is able to modify messages and signatures and thus the ECDSA algorithm becomes useless. Therefore, a group of countermeasures have used to improve security procedures. Many attacks such as signature manipulation, Bleichenbacher and restart attacks work to retrieve the private key $d$ or ephemeral $k$ generator \cite{p22}. A set of conditions (the difficulty of DPL, one-way hash, collision-resistant, $k$ unexpected) are installed to prevent such attacks on the ECDSA. We will classify attacks on the ECDSA algorithm for non-physical and physical attacks. Each category includes many attacks that try to penetrate the signatures of messages in terms of integrity, authentication, and non-repudiation. We will then explain the countermeasures applied by the researchers in their projects.
\subsubsection{Non-Physical Attacks and Security Requirements}
\begin{figure*}[t]
	\centering
		\includegraphics[width=15cm,height=8cm]{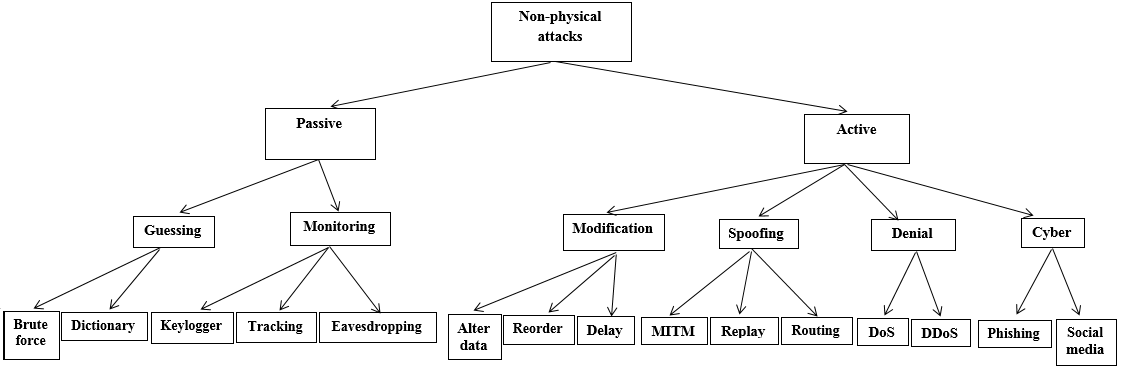}
	\caption{Classification of non-physical attacks}
	\label{survey:fig4}
\end{figure*}

These attacks do not require physical access (indirect access) to users' devices or the network to penetrate repositories' data or messages transferred between the clients and server. The attackers have used different strategies for physical attacks such as sniffing, spoofing, eavesdropping, and modification to break signatures and encoders transmitted through radio frequency signals or the Internet \cite{p143}. They implement non-physical threats to penetrate the security requirements of confidentiality, integrity, and availability (CIA) during the implementation of analysis and modification operations. Non-physical attacks can use network devices indirectly. For example, a DDoS attack uses network devices to send a stream of messages to disable the server services \cite{p144}. These attacks are categorized into passive and active attacks, according to the strategies of these attacks and the intended target of the attack.
\paragraph{Passive attacks}
These attacks have used eavesdropping, monitoring, and sniffing mechanisms to control and record the activities of network devices, and then break user authentication signatures without altering or destroying the data \cite{p145}. The attacker needs to adjust a large number of packets through the use of packet sniffing and packet analysis software to assist in the analysis and penetration of signatures. These attacks are primarily intended to penetrate authentication (signatures), encryption (confidentiality), and access control (authorisation policies). During these attacks, the attacker tries to analyze the data to gain information about messages such as sender and receiver IP addresses, transport protocol such as TCP or UDP, location, data size, and time, as this information helps him in the penetration operation \cite{p146}. Passive attacks do not perform any changes to the network data, but the attacker can use these data analyses to destroy the network in the future. This type of attack is difficult to detect by security precautions because the eavesdropper does not perform any data changes. But there is also a set of security requirements that prevent such attacks. Many research projects implement the ECDSA algorithm to prevent the passive non-physical attack. Table~\ref{survey:tab8} shows passive non-physical attacks types in many applications with ECDSA. A brief explanation of some of the most popular passive attacks as follows.
\begin{itemize}[noitemsep,nolistsep]
\item Traffic analysis and scanning (eavesdropping): Monitors and analyses data as they travel over the Internet or wireless network, for example, monitoring data size, number of connections, open ports, visitor identity and vulnerabilities in operating systems \cite{p147}.
\item Keylogger and snooping: Records authentication activities in victim devices during keystrokes monitoring of username/password \cite{p107}.
\item Tracking: Records and traces users' information such as location for network devices and personal information such as name, address, email, and age \cite{p110}.
\item Guessing: The attacker attempts to guess a user's credential by relying on words like dictionary attack or a combination of symbols (probabilities) such as brute force attack to gain access to the network \cite{p117}.
\end{itemize}
\paragraph{Active attacks}
An active attacker performs data penetration of users' identities by creating, forging, modifying, replacing, injecting, destroying, or rerouting messages as these messages move through the network's nodes or the Internet \cite{p146}. This type of attack can use sniff attack (passive attack) to collect information and then performs changes or destruction of network data. For instance, a malicious attacker intercepts money transfer messages in e-banking applications and then implements adding and deleting operations to obtain personal gains \cite{p107}. An active attacker can send false or fake signals to deceive the network nodes by linking to the fake server and then redirecting nodes’ packets to the legitimate server after modifying the data or authentication messages. Security requirements are significantly important, in particular, the authentication and integrity requirement, to prevent many attacks such as man-in-the-middle (MITM) and replay \cite{p121}. According to many research projects in \cite{p105,p148,p127,p120,p118}, ECDSA's signatures is a security solution to prevent many attacks such as modification, spoofing, denial, and cyber. Figure~\ref{survey:fig4} shows the classification of non-physical attacks. Table~\ref{survey:tab8} shows active non-physical attacks types in many applications with ECDSA. A brief explanation of some of the most popular active attacks is given in the following.
\begin{itemize}[noitemsep,nolistsep]
\item Spoofing: Many types of attacks such as MITM, replay, routing, and hijacking. The attacker intercepts and modifies credential messages to gain access to the network by MITM attack, while in replay attacks, the attacker intercepts messages and resends them later \cite{p117}. Also, routing/hijacking changes the packets paths (changing the IP address) by exploiting the vulnerability in the routing algorithms used to find the best path for moving packets \cite{p146} such as IP-spoofing and ARP-spoofing.
\item Denial: These attacks are categorized as denial of service (DoS) and distributed denial of service (DDoS). In DoS, the attacker uses his own devices to prevent the server from providing services to network members. This attack targets security requirements (CIA), while in DDoS the attacker uses his own devices as well as network devices to quickly destroy the network \cite{p110,p146,p121,p145}.
\item Cyber: An attacker creates websites such as phishing attacks or social pages for Facebook and Twitter, such as social media attacks, to trick the user into believing that these websites are legitimate websites or pages. The user enters his/her confidential information such as username, password and card number in these counterfeit websites to allow this information to be accessed by the hacker \cite{p107,p117}.
\item Modification: This attack changes, delays, rearranges users' data packets after gaining the attacker's access as a legitimate user or the attacker may be a legitimate member of the network (internal attack). Any change to this data can cause problems for the user such as changing medical or diagnostic reports in e-health applications \cite{p110}.
\end{itemize}
\paragraph{Security Requirements}
The ECDSA algorithm provides three security requirements: authentication, integrity, and non-repudiation. But integrating the ECDSA algorithm with security mechanisms such as encryption and authorisation provides many security requirements (such as confidentiality, authorisation, availability, accountability, forward secrecy, backward secrecy, auditing, scalability, completeness, anonymity, pseudonymity, and freshness) when used in applications like e-health, e-banking, e-commerce, e-vehicular, and e-governance. We will provide a brief explanation of the security requirements that provided by ECDSA as in the following.

\begin{itemize}[noitemsep,nolistsep]
\item Authentication has been used to authenticate legitimate users or data in the network to prevent anyone else from accessing it. Namely, if the users' identities or data is a trusted source in the network it is accepted, but if an unknown source it is ignored \cite{p126}. Many attacks such as brute-force, keylogger, and credential guessing attempt to penetrate the signatures' authentication service.
\item Integrity has been used to ensure that the transmitted data has not been tampered with or
edited by the adversary \cite{p145}. Many attacks such as MITM, replay, hijacking, and phishing  attempt to penetrate the signatures' integrity service.
\item Non-repudiation has been used to detect the compromised nodes via the sender who cannot deny his message \cite{p121}. Many attacks such as repudiation, masquerading, and social media attempt to penetrate the signatures' non-repudiation service.
\end{itemize}
\subsubsection{Physical Attacks and Countermeasures}
Physical attacks are passive or active attacks. The attacker applies a passive physical attack to analyse the signatures and breaks the authentication property (gets the private key) to become a legitimate user in the network. On the other hand, the active physical attack attempts to penetrate the integrity or non-repudiation property to change signatures and messages transferring between the network's nodes. Many physical attacks have applied on ECC/ECDSA. These attacks exploit some problems in these algorithms in order to access the private key \cite{p27,p49}. These problems include:
\begin{itemize}[noitemsep,nolistsep]
\item The power consumption
\item Electromagnetic radiation 
\item Computation time 
\item Errors
\end{itemize}
\paragraph{Passive attacks}
This type of attacks do not tamper with or modify the data but analyses the leaked bits of data (such as the scalar's bits).
 These attacks take advantage of the different running time for operations, power consumption and electromagnetic radiation.
 The attacker tries to get some bits leaked from $k$ to produce the full value of $k$. The attacks that exploit the power consumption and electromagnetic radiation to detect $k$ are called side channel attacks (SCAs). SCA attacks consist of a suitable model, power trace, and statistical phase \cite{p27}. In these attacks, the attacker monitors the power consumption and exploits the unintended outputs (side-channel outputs on the secret key) of the device \cite{p18}. Leakage power consumption is divided by the transition count leakage and Hamming weight, where the first depends on the state variable bits at a time, while the second depends on the number of 1-bits treated in time \cite{b4} by tracking voltages of the device. SCA uses many methods to detect $k$ bits such as the analysis of the distinction between the addition formula operations, creation of a template, statistical analysis, re-use values, special points, auxiliary registers, and the link between the register address and the key \cite{p49} (passive attack is described in Figure 3). SCA attacks (passive) consist of four major attacks:
\begin{itemize}[noitemsep,nolistsep]
\item Simple power analysis (SPA)\\
In this attack, the attacker relies on a single trace of power consumption to detect the secret key bits. The attacker extracts these bits based on power consumption discrimination in the addition formula (point addition (PA) and point doubling (PD)) \cite{p18,p49}.
\item Timing attack\\
In this attack, the attacker relies on the analysis of the execution time of the addition formula and the arithmetic operations \cite{p58}. For example, the attacker analyses the processing time for PA and PD; if the processing time is greater, it is considered PA ("1") or else PD ("0"), and repeats the process until he/she obtains all ephemeral bits $k$.
\item Template attack\\
In this attack, the attacker creates templates with a large number of traces of the controlled device. It uses multivariate normal distribution to detect the key based on power consumption during data processing. The attacker gets the ECDSA's key by matching the best template with these traces \cite{p49,p82}. 
\item Deferential power analysis (DPA)\\
In this attack, the attacker uses many traces in a statistical analysis of power consumption. The attacker uses hypothetical points in SM with stored recorded results. He/she then compares these results with the power consumption of the controlled device to obtain a valid guess in the detection of secret key bits \cite{p132}.
\end{itemize}
General countermeasures have used to prevent passive physical attacks \cite{p71}:
\begin{itemize}[noitemsep,nolistsep]
\item Elimination of relation between data and leakages.
\item Elimination of relation between fake data and real data.
\end{itemize}
The implementation of one of the former two countermeasures ensures data protection from attacks.
\begin{figure*}[t]
	\centering
		\includegraphics[width=15cm,height=6cm]{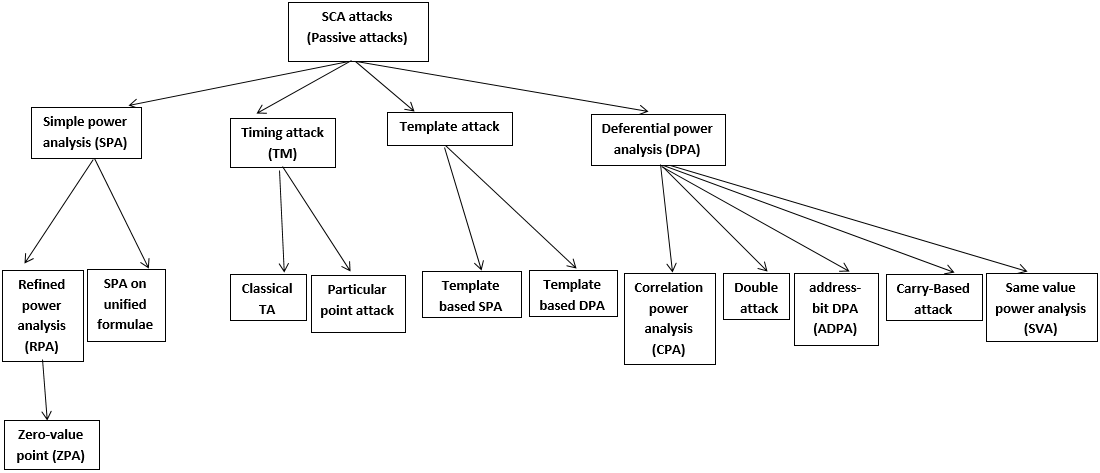}
	\caption{SCA (passive) attacks}
	\label{fig:fig2}
\end{figure*}
Jacobi form has been proposed instead of Weierstrass to prevent SCA attacks \cite{p83}. This form allows for addition formula operations to have identical time and power and this leads to prevention of SPA and DPA attacks. Unfortunately, this scheme is 70\% less efficient than the original scheme (Weierstrass). The average of field operations in their scheme is 3664 while in the original scheme it is 2136. Many recommendations have been presented with PM's endomorphisms to protect many attacks such as Pohlig-Hellman and Pollard's rho through $n \geq 2^{160}$, $ \# E(F_q)\neq q$ and $n$ is not dividable on $q^i-1$ when $1 \leq i \leq 20$ \cite{p69}. Similarly, a study presented DSA and ECDSA algorithms in detail and discussed that these algorithms become unsuitable for signing messages (integrity) if applied incorrectly, as this study has focused on the parameters validation of DSA/ECDSA to ensure strong security for these algorithms against different attacks; also, the author proved that these algorithms become strong if the parameters are well-validated \cite{p22}. In addition, \citet{p54} proposed protection mechanisms with double-base chains method against SCA attack (SPA and DPA), using side channel atomicity against SPA and randomization method against DPA \cite{p133}. Also, right-to-left NAF algorithm was proposed without precomputations and investigated security \cite{p62}. This algorithm protected against SPA, DPA, and doubling attacks through using atomicity technique countermeasures.\\
Implementation of template based SPA attack in ECDSA was investigated with microprocessor 32-bit (ARM7 architecture) \cite{p82}. The authors pointed out that this attack is applicable on ECDSA through three basics (few bits to retrieve $k$ by lattice attack, using fixed base point and microprocessors enough to build templates) to combine with their attack. Furthermore, \citet{p58} presented a scheme dependent on combining comb method with width-wNAF. But, this scheme is not resistant to SCA attacks. Therefore, they suggested using constant runtime to prevent SPA and timing (through add point of infinity (O) when bits value equal nonzero and also add it to precomputation values). Physical attacks are explained on ECC/ECDSA algorithms \cite{p49}. The authors described many SCA attacks, as they presented countermeasures and recommendations to apply ECC/ECDSA such as adding randomness, countermeasures selection, and implementation issues to prevent these attacks. Moreover, \citet{p18} analyse SPA and DPA attacks (DPA analysis neglected because it is not viable in the ECC/ECDSA algorithm but SPA is applicable to ECDSA in WSN). DPA is not applicable in the case random scalar but is applicable in the case using it against secret key $(d)$ where the attacker knows signed message \cite{p82}. SCA attack is working on the use of information leaked from the SM ($Q = kP$). They concluded that protecting the ephemeral $k$ is important in ECDSA against SPA attacks.\\
Template attack with lattice attack is presented to retrieve ECDSA's private key over prime field $(F_{p160})$ \cite{p91}. The authors pointed out that endomorphism curves can be penetrated by Bleichenbacher attack through using one bit of bias. They had a secret key for ECDSA through few number of hours when using $2^{33}$ signatures, $2^{33}$ memory and $2^{37}$ time. \citet{p27} focused on the analysis of correlation power analysis (CPA) attack on ECDSA algorithm in FPGA. They did not use traditional power models (Hamming weight/distance) in the analysis of CPA attack but used another type as a countermeasure (chosen plaintext attack). They proved possible successful feasible CPA in ECDSA on FPGA in case $k^{-1}(e+rd_A)$ or $k^{-1}e+(k^{-1}d_A)r$, but also CPA attack had no impact in ECDSA in case using equation $k^{-1}e+(k^{-1}r)d_A$ (Results obtained by DISIPA platform). Also, a SPA attack was suggested for SM penetration based on conditional subtraction of a modular multiplication. Constant time of modular multiplication is a guarantor to prevent this attack from SM analysis. On the other hand, \citet{p134} used secure-SM with randomized point coordinates to prevent SPA, DPA, and ZPA. Table~\ref{survey:tab6} shows the passive physical attacks and countermeasures.
\begin{table*}[!t]
\scriptsize
\centering
\caption{Passive physical attacks and countermeasures}
\label{survey:tab6}
\begin{tabular}{|l|l|l|l|}
\hline
\rowcolor[HTML]{EFEFEF} 
Paper                                                               & Passive physical attack (s)                                                                                                  & Countermeasure (s)                                                                                                                               & Year                                                \\ \hline
\rowcolor[HTML]{FFFFFF} 
\cite{p83}                                                        & SPA/DPA                                                                                                                      & Jacobi form                                                                                                                                      & 2001                                                \\ \hline
\rowcolor[HTML]{FFFFFF} 
\cite{p69}                                                        & \begin{tabular}[c]{@{}l@{}}Pohlig-hellman, Pollard's raho\\ Semaev-Satoh-Smart, Weil pairing\\ and Tate pairing\end{tabular} & Parameters selection                                                                                                                             & 2001                                                \\ \hline
\rowcolor[HTML]{FFFFFF} 
\cite{p22}                                                        & Bleichenbacher and restart                                                                                                   & Parameters validation                                                                                                                            & 2003                                                \\ \hline
\rowcolor[HTML]{FFFFFF} 
\begin{tabular}[c]{@{}l@{}}\cite{p54}\\ \cite{p62}\\ \cite{p133}\end{tabular} & SPA/DPA/doubling                                                                                                             & \begin{tabular}[c]{@{}l@{}}Side channel atomicity\\ and randomization\end{tabular}                                                               & \begin{tabular}[c]{@{}l@{}}2005\\ 2008\\ 2017\end{tabular} \\ \hline
\rowcolor[HTML]{FFFFFF} 
\begin{tabular}[c]{@{}l@{}}\cite{p82}\\ \cite{p18}\end{tabular} & \begin{tabular}[c]{@{}l@{}}Template/DPA\\ SPA/DPA\end{tabular}                                                               & \begin{tabular}[c]{@{}l@{}}Random base point\\ Secure-SM\\ DPA-resistant\\ Random scalar\end{tabular}                                            & \begin{tabular}[c]{@{}l@{}}2008\\ 2013\end{tabular} \\ \hline
\rowcolor[HTML]{FFFFFF} 
\cite{p58}                                                        & SPA/timing                                                                                                                   & constant runtime                                                                                                                                 & 2012                                                \\ \hline
\rowcolor[HTML]{FFFFFF} 
\cite{p49}                                                        & \begin{tabular}[c]{@{}l@{}}SPA/DPA/CPA/template\\ RPA/ZPA\end{tabular}                                                       & \begin{tabular}[c]{@{}l@{}}Adding randomness\\ appropriate countermeasure\\ selection depend on the \\ application and environments\end{tabular} & 2012                                                \\ \hline
\rowcolor[HTML]{FFFFFF} 
\cite{p91}                                                        & Template with lattice                                                                                                        & Large prime finite                                                                                                                               & 2014                                                \\ \hline
\rowcolor[HTML]{FFFFFF} 
\cite{p27}                                                        & CPA                                                                                                                          & $k^{-1}e+(k^{-1}r)d_A$                                                                                                                           & 2015                                                \\ \hline

\rowcolor[HTML]{FFFFFF} 
\cite{p132}                                                       & SPA                                                                                                                          & \begin{tabular}[c]{@{}l@{}}Constant time modular multiplication, \\ point blinding, random Projective\end{tabular}                               & 2015                                                \\ \hline

\rowcolor[HTML]{FFFFFF} 
\cite{p134}                                                       & SPA/DPA/ZPA                                                                                                                  & \begin{tabular}[c]{@{}l@{}}Secure-SM with randomized point\\ coordinates\end{tabular}                                                            & 2017                                                \\ \hline

\end{tabular}
\end{table*}
\paragraph{Active attacks}
There is another type of SCA attack, which uses errors to reveals some bits $k$, called fault attacks. This type of attack exploits the error in the dummy operations, memory block, invalid point and curve, and the difference between right and wrong results. The attacker gets an error signature by adding one bit to the generator $k$. This type of attack is considered a serious risk to ECC/ECDSA algorithms if  no proper countermeasures have applied for applications and data transfer environment (Figure 4 shows active SCA attacks). SCA attacks (active) consist of main three attacks:
\begin{itemize}[noitemsep,nolistsep]
\item Safe-error based analysis\\
In this attack, the attacker uses faults to exploit dummy operations (used to resist SPA) and memory blocks (register or memory location). Safe-error includes two types of c safe-error and m safe-error, where the first exploits the vulnerability of the algorithm whereas the second exploits the implementation \cite{p71,p136}.
\item Weak-curve based analysis\\
In this attack, the attacker tampers with the curve parameters, where the error is injected by using an invalid point and invalid curve at a specified time for the parameters that were not correctly selected or validated. The attacker uses the errors to move SM from the strong curve to the weak and thus retrieves the private key \cite{p142,p49}.
\item Differential fault analysis (DFA)\\
In this attack, the attacker compares the difference between the correct and fault results to retrieve the scalar bits. Differential fault includes two types: classic DFA and sign change FA \cite{p49, p138}. Validation of intermediate results and random ephemeral $k$ are used to prevent classic DFA attacks while Montgomery ladder and verify the final results used to prevent sign change FA attacks \cite{p71}.
\end{itemize}
General countermeasures have used to prevent active physical attacks \cite{p71}:
\begin{itemize}[noitemsep,nolistsep]
\item Error-detection
\item Error-tolerance
\end{itemize} 
The former countermeasure allows the detection of the errors that are added by the attacker whereas the latter prevents the attacker from the discovery of scaler or private key, even after providing faults.
\begin{figure*}[t]
	\centering
		\includegraphics[width=15cm,height=6cm]{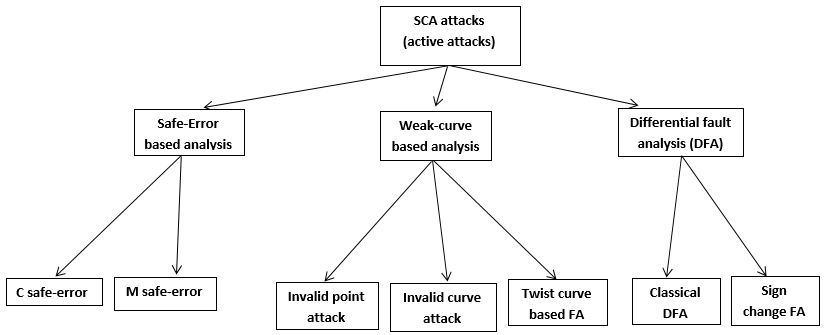}
	\caption{SCA (active) attacks}
	\label{fig:fig3}
\end{figure*}
The c safe-error attack is described on ECDSA \cite{p136}. This attack exploits dummy operations in the addition formula. The attacker can use errors with dummy operations to discover bits of $k$. In \cite{p136} they proposed the use of atomic patterns (Verneuil and Rondepierre patterns) instead of the dummy operations to repeal the c safe-error. Also, \citet{p138} suggested using a residual number system (RNS) and leak resistant arithmetic (LRA) to prevent both differential faults (comparing the difference between correct and fault results) and m safe-error (using block memory errors) that are intended to retrieve bits from ephemeral $k$.\\
An invalid point attack is carried out on a weak single curve during application skip of faults to move the points to weak Weierstrass curves to retrieve the private key \cite{p142}. The authors used point validation as a countermeasure to prevent this attack. Since the device tests the validation of the point on the curve, if it is incorrect, it generates an error message. Moreover, \citet{p137} proposed using Edwards curves with ECDSA ($F_{P128,192,256}$) instead of Weierstrass curves to prevent weak-curve attacks. They pointed out that Edwards curves provide constant time and exception-free SM so these attacks are not applicable. Similarly, \citet{p100} showed that there were safe and unsafe curves. They demonstrated that curve25519 curve with 256-bit key length is secure and fast.\\
The comprehensive survey presented information about fault attacks on symmetric and asymmetric cryptography algorithms \cite{p73}. The authors presented countermeasures to this attack in different cryptography algorithms. They pointed out that this attack may apply on ECDSA, and also presented countermeasures to prevent this attack on ECDSA. Their countermeasures used against fault attacks are the addition of the cyclic redundancy check (CRC) in the private key or using the public key to verify the signature before sending it, but this process is expensive \cite{p73}. In addition, fault attacks and types are discussed in \cite{p49,p139}. The authors have referred to countermeasures to prevent these attacks such as point validation and curve integrity check. According to \cite{p140}, many countermeasures known to repel various fault attacks such as algorithm restructuring, physical protection of the device, random techniques in computation processes and power consumption with independent implementation. However, the authors noted that fault detection, intrusion detection, algorithmic resistance and correction techniques are capable of eliminating injection fault attacks. Furthermore, \citet{p141} have provided a fault attack to retrieve the private key. Their attack depended on the injection of errors in the implementation of modular arithmetic operations in ECDSA signatures. They used multiprecision multiplication faults with scalar ($k^{-1}e + (k^{-2}r)kd $) to prevent the penetration of signatures. Recently, \citet{p135} discussed invasive SCA attacks such as fault attacks with ECC algorithms ($F_{2^m}$). The attacker injects errors into the victim's device using tools such as voltage glitch and laser. The authors pointed out that time and location greatly affect the success of fault attacks. Therefore, they proposed one multiplication module and one division module to fix the timing of all operations. Table~\ref{survey:tab7} shows the active physical attacks and countermeasures.
\begin{table*}[!t]
\scriptsize
\centering
\caption{Active physical attacks and countermeasures}
\label{survey:tab7}
\begin{tabular}{|l|l|l|l|}
\hline
\rowcolor[HTML]{EFEFEF} 
Paper                                                                & Active physical attack (s)                                                    & Countermeasure (s)                                                                                                              & Year     \\ \hline
\rowcolor[HTML]{FFFFFF} 
\cite{p73}                                                         & Fault                                                                         & CRC and public key                                                                                                              & 2004     \\ \hline
\rowcolor[HTML]{FFFFFF} 
\begin{tabular}[c]{@{}l@{}}\cite{p139}\\ \cite{p49}\end{tabular} & Fault attacks and types                                                       & \begin{tabular}[c]{@{}l@{}}Point validation, curve integrity check, \\ coherence check and combined curve \\ check\end{tabular} & \begin{tabular}[c]{@{}l@{}}2011\\ 2012\end{tabular} \\ \hline
\rowcolor[HTML]{FFFFFF} 
\cite{p140}                                                        & Fault                                                                         & \begin{tabular}[c]{@{}l@{}}Fault detection, intrusion,\\ algorithmic resistance and\\ correction techniques\end{tabular}        & 2012     \\ \hline
\rowcolor[HTML]{FFFFFF} 
\cite{p141}                                                        & Invalid point                                                                 & Point validation                                                                                                                & 2015     \\ \hline
\rowcolor[HTML]{FFFFFF} 
\cite{p136}                                                        & C safe-error                                                                  & Atomic pattern                                                                                                                  & 2016     \\ \hline
\rowcolor[HTML]{FFFFFF} 
\cite{p137}                                                        & Weak-curve                                                                    & Edwards curve                                                                                                                   & 2016     \\ \hline
\rowcolor[HTML]{FFFFFF} 
\cite{p138}                                                        & \begin{tabular}[c]{@{}l@{}}Differential fault and\\ m safe-error\end{tabular} & \begin{tabular}[c]{@{}l@{}}RNS and LRA\\ (random base permutation)\end{tabular}                                                 & 2016     \\ \hline
\rowcolor[HTML]{FFFFFF} 
\cite{p141}                                                        & Fault                                                                         & \begin{tabular}[c]{@{}l@{}}Multiprecision multiplication \\ operations\end{tabular}                                             & 2016     \\ \hline
\rowcolor[HTML]{FFFFFF} 
\cite{p135}                                                        & invasive fault                                                                & Fix computing time                                                                                                              & 2017     \\ \hline
\end{tabular}
\end{table*}
\subsection{Security Improvement Via the Random}
One of the problems that leads to weak security in the ECDSA algorithm is that there is not sufficiently random and thus ECDSA produces unsafe keys. The random increase in ECDSA prevents penetration of signed messages, whether in physical attacks or during transfer messages from the source to the destination. For instance, \citet{p3} pointed out that ECDSA used in Bitcoin suffers from poor randomness in the signature, and this leads to penetration of clients' accounts by attackers. Therefore, the keys should not be duplicated, and $k$ should not be used for more than one message. Also, the randomness source in ECDSA is extremely important to prevent key leakage \cite{b3}. If the randomness source is bad this leads to the generation of a bad signature. A bad signature allows the attacker to leak bits of the key and thus discover the private key. \citet{p92,p71,p49} pointed out to many countermeasures using randomization in scalar (blinding, splitting, and group), point (blinding and coordinates), register address, EC, and field isomorphism. This section explains some researchers' ideas to improve randomness in ECDSA algorithm.
\subsubsection{Random of Scalar}
Point multiplication $(kP)$ in the ECDSA algorithm uses a random scalar to provide strong signatures as a countermeasure against modifying the data. But the use of scalar $k$ with weak random allows the attacker to reveal $k$ and then forge signatures. The scalar should be subject to several tests to prevent weak random and make it difficult to predict scalar by the attacks.  Many methods have used to generate random scalars, such as the use of hash functions or the use of generators.\\
\textbf{Randomization source}, \citet{p78} used a SHA-1 algorithm to generate ephemeral $k$ and to gain appropriate randomness as a countermeasure against side channel attacks, where the generation of random $k$ is according to FIPS's standard. They pointed out that using a hash function with $k$ increases signature security due to changing $k$ in each message. Also, in the signature algorithm, they used equation $s=k^{-1}e+(k^{-1}r)d$ instead of $s=k^-1(e+rd)$ to prevent DPA attacks. Moreover, \citet{p23} used W7 algorithm to increase random integer $k$ in the ECDSA algorithm, which showed that the W7 generator is better than other generators in performance and area. Also, \cite{p3} focused on the weaknesses of ECC/ECDSA through the study of four SSH, TLS, Bitcoin and the Austrian citizen card protocols. They discovered that the ECDSA suffers from security weakness in the same way as previous security systems. They focused on the three points: deployment, weak keys, and vulnerable signatures. They gathered databases belonging to the four protocols and found that several cases of repeated public keys have used in SSH and TLS. Bitcoins suffer from several signatures that share nonces that allow an attacker to compute the corresponding private keys and steal coins. The Austrian citizen card as well suffers from a multi-use of the keys. The big problem with the signature is insufficient randomness in generating keys or nonces in digital signatures. For example, Debian OpenSSL. Koblitz and Menezes \cite{p81} discussed random oracle model with ECDSA in the real world. They pointed out this model is less dependent on random generators of weakness and gives more flexibility with modified ECDSA. As they explained, modified ECDSA algorithm prevents chosen-message attacks.\\ 
\textbf{Randomization techniques}, randomize with NAF was proposed to prevent DPA attack \cite{p89}. This scheme required $n/2$ for a number of point addition and $n+1$ for point doubling. The authors also proposed using randomized signed-scalar with addition-subtraction multiplication algorithm to prevent SPA attacks. Their algorithm required $n+1$ for both point addition and doubling. Their scheme is resistant to timing attacks because scalar changed in each running time depending on randomized signed-scalar representation. They explained through experimental results that their scheme is resistant to power attacks. In addition, \citet{p162} investigated the evaluation of randomization techniques such as scalar blinding and splitting with prime-256, brainpool256rl, and Ed25519 curves in ECC/ECDSA. Their results demonstrated that these techniques prevent SCA attacks. However, \citet{p160} investigated the application of scalar blinding and splitting techniques in ECC/ECDSA with asynchronous samples. They noted that these techniques are vulnerable to branch misprediction, DPA, and template attacks. To eliminate such attacks, they recommended applying execution independence for scalar $k$ or implementing parallel random branching executions. \citet{p133} noted that the integration of scalar randomization and side channel atomicity techniques is a countermeasure against SCA attacks such as SPA and DPA. Similarly, \citet{p161} pointed out that scalar blinding is a countermeasure against SPA. 
\subsubsection{Random of point representation}
The random to represent the point prevents an adversary from knowing bits in the point after its representation in binary \cite{b4}. Several methods have implemented to represent a base point at random, such as point blinding (add a random point to SM such as $kP+R$), random coordinate systems (such as Projective and Jacobian coordinates) and the use of random isomorphism.\\
\citet{p84} used randomness for intermediate values and special points through using randomized linearly-transformed coordinates (RLC) to prevent refined power analysis (RPA) and zero-value point (ZPA) attacks. They proposed using randomized initial point method (RIP) to prevent SPA and DPA attacks where RIP uses random point $(R)$ with point multiplication $(kP)$. ESDSA is vulnerable to fault attack. \citet{p35} presented a fault attack on ECDSA, attacker modifies program flow and tries to benefit from error result to get parts (bits) for ephemeral key $k$, then performs lattice attack to obtain the private key ($d$). The authors implemented this attack on point multiplication through Montgomery ladder and D\&A algorithms. The success of this attack is probability, as this attack tries to get on bits from $k$ through many signatures. They used the Jacobian coordinate system through implementation because it is efficient. Also, the authors used countermeasure to fault attack through using a discrete algorithm $l$ in point representation ($Q=(X:Y:Z;l)$, where modifying by the adversary for point addition and point doubling is detection. This scheme can prevent fault and SCA attack (with random mask). Using randomness point is recommended to prevent DPA attacks \cite{p54,p58}. \citet{p163} adopted a randomized point in PM to protect ECDSA signatures. They pointed out that a randomized point application with double-and-add-always prevents SM's bits from being attacked against multiple attacks such as ZPA, SPA, and DPA. Also, point randomization with curve Curve448 in ECC/ECDSA applied to transport layer security (TLS) \cite{p164}. The results of this research indicate that this protocol offers a 224-bit security level better than applying the Curve25519 curve with 127-bit randomization technology against SCA attacks.\\
Furthermore, \citet{p82} described countermeasure to prevent template based SPA attack using randomness coordinates in DPA resistance. Also, using verification batches reduces the total time of ECDSA verification but these batches suffer from some attacks. Randomizers used symbolic computation in ECDSA batch verification to improve security level against attacks \cite{p76}. The authors used D\&A and Montgomery ladder algorithms in their scheme. They noted that using numeric computation with randomizers in x-coordinate (D\&A) is faster and more secure than using Montgomery ladder algorithm in their scheme but original D\&A is faster than their scheme. Countermeasures (randomness coordinate) were used to prevent implementation attacks \cite{p166,p84,p78,p77,p165}. A randomized Projective coordinate (RPC) was used as a countermeasure to prevent differential power analysis (DPA) and template based SPA attacks. RPC has implemented through a random application in the $Z$ coordinate to represent the intermediate points in PM.
\subsection{Security Improvement Via PM Methods}
In Section 4, we have discussed how the PM method is used to improve ECDSA efficiency. In this Subsection, we will consider how the PM method is used to improve security in ECDSA. Improvement of the efficiency of the PM's methods is dramatically important for ECDSA algorithm, particularly when used in restricted-source environments, but at the same time maintaining the security of this algorithm is more important because the basic mission of this algorithm is security through integration of messages and protection from the modification, as well as authentication of messages and users in the network. Many attacks on the security of this algorithm have applied. Therefore, many ideas are presented by researchers to improve the security of PM's methods. PM's method such as D\&A, Window, NAF, Comb, and Montgomery is proposed to improve $kP$ through reducing the number of operations of addition and doubling. The attackers are taking advantage of these operations (such as the computation time in addition operation is longer than doubling operation) to obtain a secret key or ephemeral key. Therefore, this Section describes some of the recently improved methods for PM's security.\\
\citet{p83} used combining point blinding and randomized signed window method as a countermeasure against SPA and DPA attacks. Also, an improvement on left-to-right Window method was proposed with standard curves such as NIST and SECG as a countermeasure against SCA attacks \cite{p86}. The author added a special signed digit $(-2^w)$ to $\{1,2,...,2^w-1\}$ in scalar representation in order to get addition formula operations in the uniform formula instead of using dummy addition that were revealed by attackers, thus preventing SCA attacks. In addition, \citet{p90} improved on the idea in \cite{p86} through using right-to-left Window method with $2^w$-ary. This scheme avoided using fixed table (used in \cite{p86}) through randomization Projective, where this table allows attackers to analyze information statistically. They used two-processor in their scheme to improve the performance of PM through parallel. Moreover, window and Montgomery algorithms were improved to prevent SPA and DPA attacks \cite{p85}. The small multiple was added to ephemeral key with the idea of \cite{p86} in window method and combining addition and doubling operations into the uniform formula in Montgomery method where original Montgomery compute separately addition and doubling operations. Improvement on fixed-base comb method was presented against SCA attacks \cite{p84}. Furthermore, the authors used sequence non-zero of bit-strings (i.e. removing all zeros only using 1 and -1) to represent the ephemeral key. They pointed out that their scheme costs a little more computation time than the original comb method.\\ 
\citet{p82} used a double-and-add-always algorithm to prevent SPA attacks in their scheme. \citet{p78} used Montgomery ladder algorithm to improve PM's performance and countermeasure against SPA attacks in their scheme. \citet{p18} carried out SPA on many methods for PM such as double-and-add, window, and comb. They found comb method is optimal for their application to prevent SPA attacks. Their results explained that the Secure-CM was both more efficient timely (1.57s) and secure in its implementations. \citet{p77}  used countermeasures to prevent implementation attacks. They used zeroless signed digit (ZSD) with comb method to prevent simple power analysis (SPA).\\
\citet{p168} merged the L{\'o}pez-Dahab method with the Montgomery ladder. They pointed out that their scheme provides security against FA attacks by using the Montgomery method for coherent check and high probability detecting injected errors. \citet{p167} have shown that the Montgomery ladder with Galbraith-Lin-Scott (GLS) curves offers a 128-bit security level against SCA attacks. In 2017, \citet{p169} have noted that the ECC/ECDSA (256-bit) Comb method has a bit leakage by a SPA attack. They improved the Comb method by adding a signed representation to ephemeral $k$ to prevent SPA attacks completely.
\titlespacing*{\section}
{0pt}{6.0pt plus 0pt minus 0pt}{6pt plus 0pt}
\titlespacing*{\subsection}
{12pt}{6pt plus 6pt  minus 6pt}{3pt plus 6pt  minus 6pt}
\titlespacing*{\subsubsection}
{12pt}{6pt plus 6pt  minus 6pt}{3pt plus 6pt  minus 6pt}
\titlespacing*{\paragraph}
{12pt}{6pt plus 6pt  minus 6pt}{3pt plus 6pt  minus 6pt}
\section{ECDSA Applications} 
Recently, many applications have become available that store and move sensitive data for users between network nodes and server. These applications require powerful encryption and signature algorithms (such as ECC/ECDSA) to protect users' information. We will show the most important applications that use the ECDSA algorithm to provide security and privacy requirements for this data. Table~\ref{survey:tab8} shows ECDSA and recent applications with non-physical attacks.
\begin{table*}[!h]
\scriptsize
\centering
\caption{ECDSA and applications against passive and active non-physical attacks}
\label{survey:tab8}
\begin{tabular}{|l|l|l|l|l|}
\hline
\rowcolor[HTML]{EFEFEF} 
Paper and year                                                                  & \begin{tabular}[c]{@{}l@{}}Passive non-physical \\ attack (s)\end{tabular}                                             & \begin{tabular}[c]{@{}l@{}}Active non-physical \\ attack (s)\end{tabular}                                            & \begin{tabular}[c]{@{}l@{}}Signature algorigthm\\ and key size\end{tabular}                                                                               & Application                                            \\ \hline
\cite{p148} - 2017                                                              & \multicolumn{1}{c|}{\cellcolor[HTML]{FFFFFF}-}                                                                         & \begin{tabular}[c]{@{}l@{}}Masquerade, MITM,\\ replay and target oriented\end{tabular}                               & \begin{tabular}[c]{@{}l@{}}ECC/ECDSA \\ (160-512 bit) with \\ pairing cryptography\end{tabular}                                                           & \cellcolor[HTML]{FFFFFF}                               \\ \cline{1-4} 
\cite{p110}- 2017                                                               & \begin{tabular}[c]{@{}l@{}}Tracking and \\ eavesdropping\end{tabular}                                                  & \begin{tabular}[c]{@{}l@{}}Impersonation, replay, \\ MITM and cloning\end{tabular}                                   & \begin{tabular}[c]{@{}l@{}}ECDSA, ECDH, AES \\ and Shamir's trick\end{tabular}                                                                            & \multirow{-6}{*}{\cellcolor[HTML]{FFFFFF}\begin{tabular}[c]{@{}l@{}} E-health\\ (EHR, EMR,\\PHR)\end{tabular}}     \\ \hline
\cite{p107} - 2017                                                              & \begin{tabular}[c]{@{}l@{}}Shoulder surfing, \\ keyloggers\\ password-related \\ and direct\\ observation\end{tabular} & \begin{tabular}[c]{@{}l@{}}Phishing and\\ data breach incidents\end{tabular}                                         & \begin{tabular}[c]{@{}l@{}}ECDSA (256-bit) and \\ SHA-256\end{tabular}                                                                                    & \cellcolor[HTML]{FFFFFF}                               \\ \cline{1-4} 
\cite{p117} - 2017                                                              & Wifi sniffing                                                                                                          & \begin{tabular}[c]{@{}l@{}}Phishing, impersonation, \\ MITM and reflection\end{tabular}                              & \begin{tabular}[c]{@{}l@{}}ECDSA (224-bit) and \\ ECIES\end{tabular}                                                                                      & \multirow{-10}{*}{\cellcolor[HTML]{FFFFFF}\begin{tabular}[c]{@{}l@{}}E-banking\\(OB, MB,\\CC, ATM)\end{tabular}}    \\ \hline
\cite{p105} - 2017                                                              & Stealthy                                                                                                               & Heartbleed                                                                                                           & \begin{tabular}[c]{@{}l@{}}ECDSA (256-bit) to \\ support SIGaaS\end{tabular}                                                                              & \cellcolor[HTML]{FFFFFF}                               \\ \cline{1-4}
\cite{p118} - 2017                                                              & Eavesdropping                                                                                                          & MITM                                                                                                                 & \begin{tabular}[c]{@{}l@{}}ECDSA (256-bit) with\\ threshold  signatures\end{tabular}                                                                      & \multirow{-4}{*}{\cellcolor[HTML]{FFFFFF}\begin{tabular}[c]{@{}l@{}}E-commerce\\(Bitcoin,Litcoin,\\Freicoin, Peercoin)\end{tabular}}   \\ \hline
\cite{p120} - 2017                                                              & \multicolumn{1}{c|}{\cellcolor[HTML]{FFFFFF}-}                                                                         & \begin{tabular}[c]{@{}l@{}}Fake entity and\\ position counterfeit\end{tabular}                                       & \begin{tabular}[c]{@{}l@{}}ECDSA (160,192,224,\\ 256, 384, 521) with \\ various curves\end{tabular}                                                       & \cellcolor[HTML]{FFFFFF}                               \\ \cline{1-4}
\cite{p121} - 2017                                                              & Tracking                                                                                                               & \begin{tabular}[c]{@{}l@{}}Modification, Sybil,\\ impersonation, DoS,\\ bogus information.\\ and replay\end{tabular} & \begin{tabular}[c]{@{}l@{}}ECDSA (224-bit) \\ with pseudonym\end{tabular}                                                                                 & \multirow{-8}{*}{\cellcolor[HTML]{FFFFFF}\begin{tabular}[c]{@{}l@{}}E-vehicular\\(V2V, V2I)\end{tabular}}  \\ \hline
\begin{tabular}[c]{@{}l@{}}\cite{p124} - 2015\\ \cite{p126} - 2015\end{tabular} & \multicolumn{1}{c|}{\cellcolor[HTML]{FFFFFF}-}                                                                         & \begin{tabular}[c]{@{}l@{}}Private key recovery \\ and tampering\end{tabular}                                        & \begin{tabular}[c]{@{}l@{}}Signatures algorithms\\ (ECC/ECDSA, RSA) \\ with one time public \\ key infrastructure and\\ parameters selection\end{tabular} & \cellcolor[HTML]{FFFFFF}                               \\ \cline{1-4} \cline{1-4}
\begin{tabular}[c]{@{}l@{}}\cite{p129} - 2013\\ \cite{p127} - 2016\end{tabular} & \multicolumn{1}{c|}{-}                                                                         & Cyber                                                                                                                & ECC/ECDSA                                                                                                                                                 & \multirow{-10}{*}{\cellcolor[HTML]{FFFFFF}\begin{tabular}[c]{@{}l@{}}E-governance\\(G2C, G2B,\\ G2E,G2G)\end{tabular}} \\ \hline
\end{tabular}
\end{table*}
\subsection{Electronic-Health}
E-health provides services that allow healthcare providers and patients to share medical records across various health centers, such as hospitals, clinics, and even the home. These services provide facilities to improve the health of patients. Because of efficient methods of electronically sharing patient data rather than traditional paper-based methods, patient health data is available anywhere and anytime for healthcare providers and patients. Health institutions and researchers are seeking to develop these applications to improve the quality of care, disease diagnosis, and remote medical surveillance \cite{p109}. E-health includes many systems such as electronic health record (EHR), electronic medical record (EMR) and personal health record (PHR), which are used efficiently to share medical records either globally (EHR) or locally (EMR) and are administered either by the authority provider (EHR and EMR) or by the patient (PHR) \cite{p112}. These applications also suffer from many problems such as complexity in computations, lack of resources, precision management, and mobility. However, the main problem that threatens the acceptance of these systems for patients and providers is the privacy of patients' data. This data or medical reports from WSN, laptop, and phone to an e-health server are vulnerable to attacks and intrusion because the Internet and wireless network are unsafe environments.\\Data privacy is a key issue for any e-application specifically for e-health applications. These applications require security and privacy mechanisms such as authorisation policies, encryption algorithms, and robust signatures to protect medical repositories for patients from malicious attacks. Many examples of security refer to threats implemented against e-health:
\begin{itemize}[noitemsep,nolistsep]
\item In 2013, there were penetration attacks on healthcare data in US hospitals. These attacks revealed 85.4\% of the medical records (protected health information (PHI)) of the 5 largest incidents for patients' data \cite{p149}.
\item In 2016, Apple Health (Medicaid) was exposed for data breach. This attack revealed 370,000 records for clients in Apple Health (Washington state) \cite{p158}.
\item In 2017, an unauthorised individual penetrated EHR the New Jersey Diamond Institute for Fertility and Menopause. The hacker revealed PHI to 14633 record containing patients' information such as names, birth dates, social security numbers, and sonograms \cite{p150}.
\end{itemize}
According to the Vormetic report on data security 2016, healthcare is one of the sectors that is most vulnerable to hackers' attacks and thus requires increased efforts to secure health data by 64\%. This report stated on 21 January 2016 that 91\% of enterprises suffer from vulnerability threatening data security (internal and external attacks). The study of security data included several countries such as Australia, USA and Germany \cite{p114}. Therefore, many systems such as national e-health transition authority (NEHTA) in Australia and health insurance portability and accountability act (HIPAA) in the USA recommended applying security and privacy in e-health applications correctly and accurately to prevent security threats \cite{p111}.\\
In addition, the accuracy of medical data is critical in diagnosing diseases and determining the condition of patients during their online or wireless transmission from patient to a healthcare provider. Also, penetration of the medical data with diseases such as addiction, HIV infection, sexual and dermatological conditions can lead to harassment, discrimination, even death of the patient if data reports change during the transfer from client to server \cite{p113}. These penetrations show that medical records in e-health require a high level of security and privacy. Many modern projects implement the ECDSA algorithm for high security and efficient performance. ECDSA signatures (Koblitz with the key of 163-bit) and energy-harvesting techniques were proposed to monitor patients' conditions by WSN in e-health \cite{p116}. The authors applied the patient's health surveillance scenario during the deployment of a set of sensor nodes such as TelosB, Micaz, and MagoNode ++. They generated a signature with a time of 300ms and an energy consumption of 7mj. They concluded that these results reduce time and energy in WSN to serve e-health applications that deal with massive data. \citet{p148} proposed using ECC/ECDSA to protect patients' information in collaborating clouds healthcare based on pairing based cryptography. Also, \citet{p110} designed a scheme to support the privacy and security of patients' data in e-health. This scheme applied the ECDSA's signatures in the RFID tag to provide mutual authentication in the protection of medical records. They pointed out that their scheme is capable of preventing man-in-the-middle attacks, tracking, eavesdropping, impersonation, replay, and cloning. They used the Shamir scheme to reduce the cost of PM in signature verification (ECDSA) as well as security support. Their scheme provided fewer calculations and more security in the process of authentication of medical records in e-health.
\subsection{Electronic-Banking}
E-banks have used the personal information of their customers to authorise access to their bank accounts.  Many systems have applied for dealing with bank accounts such as online banking (OB), mobile banking (MB), credit card (CC) \cite{p159} and automated teller machines (ATM) which require signature algorithms to protect confidential information for customers. E-banking implements a set of security measures to prevent hackers and Internet thieves. Hackers are trying to create a gap in these systems to hack users' accounts. They have applied high-tech to penetrate customers' credits. Therefore, the process of authentication for legitimate users in e-banking applications is extremely important. Many examples of security refer to threats implemented against e-banking applications:
\begin{itemize}[noitemsep,nolistsep]
\item In 2012, millions of clients blocked access to their accounts at the Royal Bank of Scotland \cite{p106}.
\item In 2015, the website www.000webhost.com was hacked by a malicious attack. The result was the attacker's access to 13 million user accounts; the hacked data contained personal information for customers (such as name and plaintext password) \cite{p107}.
\item In 2016, DDOS attacks were perpetrated against HSBC bank, which is one of the biggest banking names in the world; these attacks led to the suspension of service for two days and prevented customers from accessing their accounts \cite{p106}.\hfill \\ \\
\end{itemize}
\vspace*{-2.9em}
In the current era, authentication systems based on username/password are no longer safe in protecting user accounts \cite{p107}. We have investigated the adoption of ECDSA signatures in modern e-banking applications and their ability to prevent hacking threats. In 2017, \citet{p107} proposed a scheme to protect e-banking applications. They designed a protocol to authenticate customers in e-banking applications. Also, they pointed out that their scheme prevents e-banking attacks such as phishing, shoulder surfing, keyloggers, data break accidents and password-related attacks. Their scheme relied on the ECDSA (256-bit) algorithm with SHA-256 to sign the ticket. This ticket implements security mechanisms to protect users' accounts such as one-time username, session key, ticket validity period, timestamp and add account permission for each login operation. They pointed out that their scheme provides protection against previous attacks as it prevents reuse of credits. Moreover, \citet{p117} discussed that short message service (SMS) is not reliable in e-banking applications. These applications exchange these messages with their customers in a dense and daily way over the Internet or wireless network. They proposed a model for protecting bank messages by using ECDSA (p-224) and the elliptic curve integrated encryption scheme (ECIES). They integrated the ECIES algorithm with signature verification in ECDSA to support the authentication process in e-banking applications. Their model achieves three stages (authentication, transaction, and notification) to complete the authorisation process between sender and receiver. They pointed out that their scheme is efficient and reliable for real-time banking applications, against the attacks of wifi sniffing, reflection, replay, man-in-the-middle, phishing, and impersonation, and provides end-to-end communication security between the client and server. They demonstrated that the ECC/ECDSA algorithm provides high performance in e-banking applications because these algorithms use small keys and an appropriate computation cost for these applications compared to other public key algorithms (such as RSA, DSA, and DH).
\subsection{Electronic-Commerce}
The Internet is important medium for providing services, sharing information, buying and selling electronic products, applications, business transactions on an internal or external e-commerce level. Business and e-commerce transactions information have been shared between several parties, such as governments, companies, and customers (consumers) via websites. The Internet has brought a tremendous amount of positives but it has also made some negatives. The most important drawback is hacking threats to personal and sensitive information to get gains and benefits \cite{p105}. The progress of the huge development of this technology has brought with it security risks that threaten the personal information of users and attempt to detect private keys in signing and encrypting data. Due to a high demand for consumers and companies to market products online, e-commerce requires secure signature algorithms and is efficient to deal with tens of thousands of requests per second. Many examples of security refer to threats implemented against e-commerce:
\begin{itemize}[noitemsep,nolistsep]
\item In 2016, the report in \cite{p152} pointed out that many popular websites have been hacked by credential replay attacks (password reuse) for more than three billion user accounts. Fraud attacks on e-commerce in the united states are expected to hit \$7 billion by 2020.
\item In 2016, phishing attacks target e-commerce where 44\% of threaten customers and 11\% of public activity \cite{p151}.
\item In 2017, the study of IP converge data services (IPC) stated that more than 16,600 DDoS attacks were revealed on e-commerce deals particularly on Valentine's Day and Chinese New Year \cite{p153}. 
\end{itemize}
Many currency signing systems have used in e-commerce applications such as Bitcoin, Litecoin, Freicoin and Peercoin \cite{p119}. Most of these systems implement ECDSA signatures to support security features (authentication, integrity, and non-repudiation). A scheme to secure e-commerce and online transactions was proposed in \cite{p105}. The authors developed a graphic processing unit-universal elliptic curve signature server (Guess) with the adoption of the ECDSA signature mechanism (key of 256-bit binary) to provide signatures with high-performance computations. They applied signature as a service (SIGaaS) on an enterprise private cloud (EPC) to secure e-payment over the Internet. They pointed out that the Internet is an insecure medium because of the variety of attacks that try to hack private keys. Also, they described SIGaaS as impervious to private key attacks such as heartbleed and stealthy newly emerged. Their scheme described the authentication issues between tenants and Guess. Therefore, they have demonstrated that e-commerce applications require robust authentication. For instance, Chinese Alipay website fulfills huge requests and exchanges personal information for users. This website performs more than 85,900 transactions per second. Therefore, this website requires an efficient signature algorithm in terms of speed, time and security requirements to protect these transactions online. The authors stated that the graphics processing unit (GPU) with ECDSA showed high performance and security in dealing with e-commerce websites such as the games market. Furthermore, \citet{p118} used ECDSA to implement two-factor authentication with e-commerce applications (Bitcoin wallet). They implemented a mutual signature protocol with signature verification over a separate communication channel. Their scheme applied a threshold to support a set of signatures, instead of one signature as well as using the hash value in checking the payer and payee. They argued that ECDSA and threshold signatures solved Bitcoin security problems such as monitoring transaction details (block chain public) and transaction size. They also pointed out that randomization is significantly important in preventing malicious attackers from discovering the secret key in ECDSA.
\subsection{Electronic-Vehicular} 
Vehicular ad hoc network (VANET) applications are important modern applications to secure people's lives on the road. These applications are a collection of network cars that share information such as police, emergency, and smart taxi cars. These applications regulate network traffic to ensure safety for users while enforcing road laws. Several features for VANET applications have used such as self-regulation, distributed communications environment, and dynamic topology. These applications have been designed to avoid traffic congestion, prevent accidents, and prevent violations of road rules. The primary objective is to protect passengers on the road. This type of network operates with two categories: vehicle to vehicle (V2V), and vehicle to infrastructure (V2I). VANET is a class of mobile ad hoc network (MANET) network but it works regularly and not randomly. The critical issue and great challenge for this type of application is security due to the VANET network sharing nodes information in the open medium \cite{p108}. To protect users in a VANET network, encryption and signature mechanisms should be used to prevent the risks of modifying messages between nodes. Many attacks threat the VANET application such as impersonation, Sybil, modification, DDoS and reply. To prevent these attacks the confidentiality, integrity, and availability (CIA) security requirements should be applied. Many examples of security refer to threats implemented against e-vehicular:
\begin{itemize}[noitemsep,nolistsep]
\item In 2015, a hacker successfully carried out a cyber attack to hijack and control a Jeep car over the Internet. This attack prevented the driver from taking control of his car \cite{p155}.
\item In 2017, the report in \cite{p154} stated that a hacker penetrated board messages for highways in California. This hacker targeted to distract drivers and endanger them. The report reported that 58.85\% of attacks such as DDoS, cyber and brute forcing  were rated as high risk for VANET applications.
\end{itemize}
The ECDSA algorithm was used to authenticate messages with VANET \cite{p30}. Though the results, the authors noted delay generation key, signature, and verification increase with an increase in curve size. This scheme provides integrity, authentication, and non-repudiation, as it is convenient for constrained-source networks. Also, \citet{p120} have provided a security analysis of the ECDSA algorithm (key lengths of 160, 192, 224, 256, 384 and 521) with VANET (V2V) applications. They applied many curves to the ECDSA (OpenSSL library) such as prime, pseudorandom binary, Koblitz binary. They also focused on data transfer scenarios and used the Riverbed Modeler tool to analyze data transferred between vehicles. Their scheme discussed whether VANET can be exposed to many attacks such as fake entity, position counterfeit, and Sybil. To generate sufficient security in these applications, at least the NIST p-224 and NIST p-256 keys should be used in ECDSA. They noted that the important parameters that affect the efficiency of signatures in an e-vehicular environment are the size and time of signature generation and verification. They concluded that ECDSA p-224 or p-256 provides efficiency (time and size of data) and adequate security to secure e-vehicular (V2V) applications. In addition, \citet{p121} argued that the privacy of location information and identity should be safe in e-vehicular applications because changing this data could lead to fatal accidents. They applied a method of authentication based mainly on the signatures of ECDSA with a pseudonym to provide anonymity to the vehicular applications. Also, they pointed out that their scheme is capable of preventing malicious attacks such as location tracking, modification, impersonation, bogus information, Sybil, replay, and DOS. Their scheme supports standard security requirements as well as traceability and low overheads. In their scheme, the pseudonym consists of signing ECDSA, timestamp and a random number. They pointed out that a lightweight pseudonym is sufficient to support security in e-vehicular applications against malicious attacks.\hfill
\titlespacing*{\section}
{0pt}{6.0pt plus 0pt minus 0pt}{6pt plus 0pt}
\titlespacing*{\subsection}
{12pt}{6pt plus 6pt  minus 6pt}{3pt plus 6pt  minus 6pt}
\titlespacing*{\subsubsection}
{12pt}{6pt plus 6pt  minus 6pt}{3pt plus 6pt  minus 6pt}
\titlespacing*{\paragraph}
{12pt}{6pt plus 6pt  minus 6pt}{3pt plus 6pt  minus 6pt}
\vspace*{0.3em}
\subsection{Electronic-Governance}
E-governance is an application that provides services to stakeholders through various communication technologies such as Internet, SMS, and email to simplify the procedures of services for citizens, business, and government agencies. This application consists of four types of transactions involving government to citizens (G2C), government to business (G2B), government to employee (G2E), and government to government (G2G) \cite{p125} to deal with various data types of stakeholders. The adoption of these applications in society should meet a high level of citizen confidence through a range of requirements such as flexibility, transparency, accountability, manageability, good governance, as well as trustworthy (security and privacy) services \cite{p123}. According to a United Nations report, a significant increase is in the development of e-governance applications in some countries of the world, but the issue that restricts their use is the security of data and the privacy of the stakeholders (citizen, government, business, employees, intermediaries and service providers) against hacking attacks \cite{p122}. These applications are a large pool of databases of diverse sources; therefore, these applications are exposed many risks such as legitimate users, e-government employees, foreign intelligence agencies, business enterprises, terrorist organizations, software providers, and natural disasters. \citet{p125} referred to several attacks against e-governance such as DOS, insider, impersonation, replay, and offline password guessing. Many examples of security refer to threats implemented against e-governance:
\vspace{-0.59em}
\begin{itemize}
\itemsep -0.3em 
\item In 2012, 112 Indian government websites such as the Planning Commission and the Finance Ministry were hacked. The hacker stopped these websites from working for weeks \cite{p156}. 
\item In 2017, a cyber attack was carried out against Australian government websites such as the Finance Department and Australian Electoral Commission. This attack revealed online sensitive information of more than 50,000 records \cite{p157}.
\end{itemize}
\vspace{-0.8em}
To secure citizen's data in e-governance applications, security requirements should be applied. Several procedures have used to secure the privacy of users such as server security, network security, data security, workstation security as well as physical and environmental security \cite{p122}. Most developed countries such as USA, UK, and some other NATO member countries use ECC/ECDSA to secure the privacy of citizens' data, as well as to protect interoperability between governments \cite{p128} and internal communications to prevent data fraud attacks. \citet{p124} have demonstrated the importance of applying public signatures to e-governance that authenticate stakeholders in the system. The availability of efficient digital signatures will provide a powerful e-authentication mechanism that will allow citizens to safely handle government documents (such as property tax, income tax, license, visa, passport, and identity card services). These applications invest technologies such as cloud computing and Big Data in dealing with large amounts of citizen data, but also provide significant challenges for developers of these technologies in terms of security and privacy. Moreover, \citet{p126} argued that e-governance applications store large amounts of citizens' data in various technologies such as cloud computing. They proposed applying ECDSA signatures to protect cloud computing data from change with recommendations for using ECDSA with accurate parameters. In addition, \citet{p127,p129} have noted that ECC/ECDSA is the best algorithm for signing and encrypting sensitive data transmitted through e-governance applications. They also discussed that several SCA attacks try to penetrate the private key in ECDSA. Furthermore, they explained that there is no countermeasure to block all hacker attacks, but a set of countermeasures should be used to secure data sharing.
\section{Conclusion and Future Work} 
In this paper, we made a detailed study of the security and efficiency of the ECDSA algorithm. This study enabled us to see the recent improvements for ECDSA in terms of security, efficiency, and application. This updated improvement may possibly inspire some ideas to improve some of the methods in the security and efficiency of the ECDSA. We hope that this study will be useful for researchers to locate research ideas in securing and improving the efficiency of the ECDSA algorithm, especially when used in the constrained source devices. Several future works that we intend to accomplish:
\begin{itemize}[noitemsep,nolistsep]
\item We intend to implement the ECDSA algorithm in e-health applications to apply the integrity and authentication properties to protect patients' data by the design of the authentication and authorisation protocols for users in the network.
\item We intend to implement lightweight curves such as Edward curve, the Montgomery SM method, and $\lambda$-Projective in \cite{p79} that are critical procedures to increase the efficiency of implementing ECDSA signatures in accessing a massive database of e-health.
\item We intend to use lightweight hash algorithms to generate random ephemeral $k$ and personal information for healthcare users. These algorithms provide high performance in computations operations.
\item We intend to investigate the application of the ECDSA algorithm with homomorphic and anonymity mechanisms to secure source-constrained devices data such as WSN. These devices require efficient and secure mechanisms to protect patients' data when transferred from WSN to e-health.
\end{itemize}

\footnotesize{
\paragraph*{Acknowledgements}
We would like to acknowledge and thank the efforts of Dr. Barbara Harmes, and Hawa Bahedh as well as the valuable feedback of the reviewers.

\nocite{}
\bibliographystyle{IEEEtranN} 
\bibliography{ref}
\end{document}